\documentclass[useAMS,usenatbib,usegraphicx]{mn2e}
\usepackage{times}
\usepackage{amssymb}

\title[Flexion in COSMOS]{Probing Galaxy Dark Matter Haloes in COSMOS with Weak Lensing Flexion}
\author[M.~Velander, K.~Kuijken \& T.~Schrabback]{Malin Velander\unskip$^{1}$\thanks{E-mail: velander@strw.leidenuniv.nl}, Konrad Kuijken\unskip$^{1}$ and Tim Schrabback\unskip$^{1}$ \\
$^{1}$Leiden Observatory, Leiden University, PO Box 9513, 2300 RA Leiden, the Netherlands}

\begin{document}
\date{Accepted 2010 November 24. Received 2010 November 24; in original form 2010 November 12}
\pagerange{\pageref{firstpage}--\pageref{lastpage}} \pubyear{2010}
\maketitle

\label{firstpage}

\defcitealias{shj10}{S10}
\defcitealias{kui06}{KK06}
\defcitealias{bgr06}{B06}

\begin{abstract}
Current theories of structure formation predict specific density profiles of galaxy dark matter haloes, and with weak gravitational lensing we can probe these profiles on several scales. On small scales, higher-order shape distortions known as flexion add significant detail to the weak lensing measurements. We present here the first detection of a galaxy-galaxy flexion signal in space-based data, obtained using a new Shapelets pipeline introduced here. We combine this higher-order lensing signal with shear to constrain the average density profile of the galaxy lenses in the Hubble Space Telescope COSMOS survey. We also show that light from nearby bright objects can significantly affect flexion measurements. After correcting for the influence of lens light, we show that the inclusion of flexion provides tighter constraints on density profiles than does shear alone. Finally we find an average density profile consistent with an isothermal sphere. 
\end{abstract}


\section{Introduction}
Weak gravitational lensing is a powerful technique for studying the distribution of matter in the universe due to its ability to model the matter distribution in foreground structures, independent of the nature of the matter present. As the light from background sources is bent around foreground lenses, the galaxy images get distorted by the tidal gravitational field. The first-order distortion is known as shear and is essentially an elongation of the image causing the source galaxy to appear stretched in one direction. This type of distortion measurement has been used in a wide variety of cosmological studies ranging from modeling the large-scale structure using cosmic shear \citep[see e.g.][for reviews]{wae03,hoe08,mvw08} to determining galaxy halo shapes using galaxy-galaxy lensing \citep{hyg04,mhb06,phh07}.

First described by \citet{gol02}, the second-order distortion is a relatively new addition which has since been named flexion \citep{gol05,bgr06}. There are two types of flexion relevant to weak lensing studies: the first flexion induces a skewness of the brightness profile whilst the second flexion is a three-pronged distortion. In combination with shear, these distortions cause the well-known banana shape of lensed source images. As flexion is effectively the gradient of shear, it is sensitive on small scales. This makes it an important complement to shear which is sensitive on relatively large scales only. By virtue of this, and of the orthogonality of the three measurements, flexion is highly beneficial to investigations of the inner profiles of dark matter haloes, where baryons become important, and to the detection of substructure in cluster haloes. Indeed, it was recently shown \citep{els10} that mass reconstructions profit from the use of flexions in combination with shear, and flexion has already been used to constrain the halo mass distribution and to detect substructure in clusters of galaxies \citep{lkg10,ouf08}. To provide more information on substructure and mass profiles, there are currently new statistical flexion tools being developed \citep[eg.][]{lkw09,leo10,bar10}. Another application, as discussed in \citet{haw09}, is to use both flexions in combination with shear to significantly tighten the constraints on galaxy halo ellipticities compared to using shear alone.  

The shape measurement technique known as Shapelets \citep{ref03a,ref03b} works by decomposing a galaxy image into a series of 2D Hermite polynomials. These provide a simple framework for describing the main galaxy image distortion operators, such as shear and flexion, and the convolution with the point-spread function (PSF). Due to the flexible treatment of the PSF, the Shapelets formalism has an advantage over the currently most widely used shape measurement method, KSB \citep*[from][]{ksb95}, since KSB uses an idealised model for the PSF whilst Shapelets is more versatile. The KSB equivalent for flexion is known as HOLICs \citep{ouf07}.

Since the field of weak lensing is relatively new, lensing measurements are continuously being improved in accuracy and applicability. Being a statistical technique, however, the accuracy of the weak lensing results depends heavily on the amount of data available. Galaxy-galaxy flexion has been tentatively observed \citep{gol05} using the ground-based Deep Lens Survey (DLS), but to further investigate galaxy-size haloes more and better data is needed. With large surveys such as the Canada-France-Hawaii Telescope Legacy Survey (CFHTLS) and the Red Sequence Cluster Surveys (RCS, RCS2) available, and new surveys like the 1500 square degree Kilo-Degree Survey (KiDS) imminent, the future looks bright. However, a space-based data set provides better resolution and such a data set is already accessible to us: the HST COSMOS survey. Using this data we will in this paper improve on the galaxy-galaxy flexion measurements of \citet{gol05}.

This paper is organised as follows: in Section~\ref{sec:flexion} we review the formalism for shear and flexion, whilst we review the Shapelets method in Section~\ref{sec:shapelets} with a description of our implementation (dubbed the MV pipeline) in Section~\ref{sec:shapelets_mv}. In Section~\ref{sec:tests} we test the MV pipeline on simulations and in Section~\ref{sec:cosmos} the pipeline is applied to data from the COSMOS survey. We conclude in Section~\ref{sec:discussion}.

Throughout this paper we assume the following cosmology \citep[WMAP7;][]{ksd10}:

$(\Omega_{M},\Omega_{\Lambda},h,\sigma_{8},w) = (0.27,0.73,0.70,0.81,-1)$

\section{Shear and Flexion}\label{sec:flexion}
We begin by briefly reviewing the weak lensing formalism. Flexion is a second-order lensing effect first introduced by \citet{gol05} and further developed by \citet{bgr06} \citepalias[hereafter][]{bgr06}. It arises  from the fact that convergence and shear are not constant across a source image, and can be used to describe how these fields fluctuate. In the weak lensing regime, the lensed surface brightness of a source galaxy, $f(\mathbf{x})$, is related to the unlensed surface brightness, $f_0(\mathbf{x})$, via
\begin{equation}
f(\mathbf{x}) \simeq \left\{1+\left[(A-I)_{ij}x_j + \frac{1}{2}D_{ijk}x_jx_k\right]\frac{\partial}{\partial x_i}\right\}f_0(\mathbf{x}).
\end{equation}
Here $I$ is the identity matrix, $x_i$ denotes lensed coordinates, and
\begin{equation}
\mathbf{\mathrm{A}} = \left(\begin{array}{cc} 1-\kappa-\gamma_1 & -\gamma_2 \\ -\gamma_2 & 1-\kappa+\gamma_1\end{array}\right)
\end{equation}
with \mbox{$\kappa=\frac{1}{2}(\psi_{xx}+\psi_{yy})$} a second derivative of the lensing \mbox{potential $\psi$}, where subscripts denote partial differentiation. \mbox{$\gamma_1=\frac{1}{2}(\psi_{xx}-\psi_{yy})$} and \mbox{$\gamma_2=\psi_{xy}$} are the two components of the complex shear \mbox{$\gamma = \gamma_1 + i\gamma_2$}. The matrix
\begin{equation}
D_{ijk} = \frac{\partial A_{ij}}{\partial x_k}
\end{equation}
describes how convergence and shear vary across a source image. We can re-express $D_{ijk}$ as the sum of two flexions: \mbox{$D_{ijk} = \mathcal{F}_{ijk} + \mathcal{G}_{ijk}$}. The two flexions, the first flexion $\mathcal{F}$ (known as \mbox{F flexion} or one-flexion) and the second flexion $\mathcal{G}$ (known as \mbox{G flexion} or three-flexion), are the derivatives of the convergence and shear fields. There are four flexion components, each of which may be written in terms of the third derivatives of the lensing potential \citep{haw09}:
\begin{eqnarray}
\mathcal{F}_1 &=& \frac{1}{2}(\psi_{xxx}+\psi_{yyx}) \\
\mathcal{F}_2 &=& \frac{1}{2}(\psi_{xxy}+\psi_{yyy}) \\
\mathcal{G}_1 &=& \frac{1}{2}(\psi_{xxx}-3\psi_{xyy}) \\
\mathcal{G}_2 &=& \frac{1}{2}(3\psi_{xxy}-\psi_{yyy})
\end{eqnarray}
where \mbox{$\mathcal{F} = \mathcal{F}_1 + i\mathcal{F}_2$} and \mbox{$\mathcal{G} = \mathcal{G}_1 + i\mathcal{G}_2$} are the complex F and G flexions respectively. The full matrices $\mathcal{F}_{ijk}$ and $\mathcal{G}_{ijk}$ in terms of the four flexion components are written explicitly in \citetalias{bgr06}.

\section{Shapelets}\label{sec:shapelets}
The Shapelets basis function set was introduced by \citet{ref03a} and is more fully described there. In summary, the surface brightness of an object $f(\mathbf{x})$ can be expressed as a sum of orthogonal 2D functions
\begin{equation}
f(\mathbf{x}) = \sum_{a=0}^{\infty}\sum_{b=0}^{\infty}s_{ab}B_{ab}(\mathbf{x};\beta)
\end{equation}
where $s_{ab}$ are the Shapelets coefficients and the Shapelets basis functions $B_{ab}(\mathbf{x};\beta)$ are defined as
\begin{equation}
B_{ab}(\mathbf{x};\beta) = k_{ab}\beta^{-1}e^{-\frac{|\mathbf{x}|^2}{2\beta^2}}H_a(x/\beta)H_b(y/\beta).
\end{equation}
Here $k_{ab}$ is a normalization constant, $\beta$ is the Shapelets scale radius, $(x,y)$ are coordinates on the image plane and $H_n(x)$ is a Hermite polynomial of order $n$. The Shapelets basis functions are easily recognised as the energy eigenstates of the 2D Quantum Harmonic Oscillator (QHO). The formalism developed for the QHO can also be applied to Shapelets, providing analytical expressions for transformations such as shear and flexion. In theory, an object can be perfectly described through a decomposition into Shapelets up to order \mbox{$n\rightarrow\infty$} but in practice the expansion has to be truncated. We truncate at combined order \mbox{$n_\mathrm{max} = a + b$} to avoid introducing a preferred direction.

Convolution with the point-spread function (PSF) can also be done analytically in the Shapelets formalism by simply multiplying the Shapelets expansion by a PSF matrix $\mathbf{P}$:
\begin{equation}
\mathbf{P}_{a_1a_2b_1b_2}(\beta_{\mathrm{obj}},\beta_{\mathrm{con}}) = \sum_{a_3,b_3}C_{a_1a_2a_3}^{\beta_{\mathrm{con}}\beta_{\mathrm{obj}}\beta_{\mathrm{psf}}}C_{b_1b_2b_3}^{\beta_{\mathrm{con}}\beta_{\mathrm{obj}}\beta_{\mathrm{psf}}}p_{a_3b_3}
\end{equation}
where $p_{ab}$ are the Shapelets coefficients of the PSF and $\beta_{\mathrm{psf}}$, $\beta_{\mathrm{obj}}$ and $\beta_{\mathrm{con}}$ are the scale radii of the PSF, the object and the resulting PSF convolved object respectively. $C_{nml}^{\beta_1\beta_2\beta_3}$ is a convolution tensor which depends on the different scale radii and the full expression is given in \citet{ref03a}.

\subsection{The MV Pipeline}\label{sec:shapelets_mv}
\begin{figure}
\includegraphics[width=75mm,angle=270]{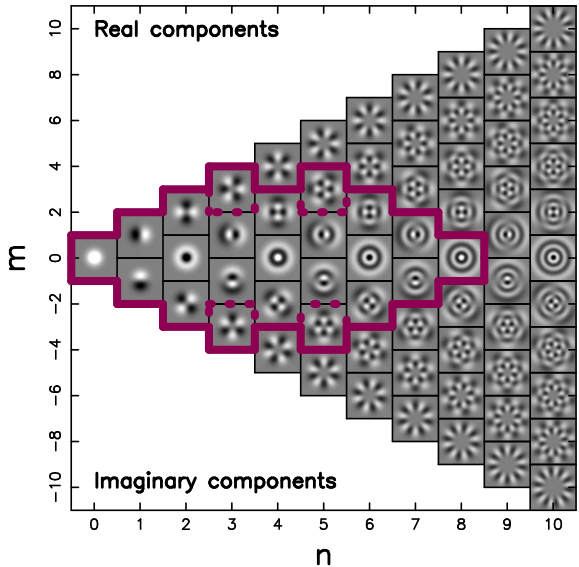}
\caption{ Polar Shapelets basis functions up to a maximum Shapelets order of \mbox{$n_\mathrm{max}=10$}. For \mbox{$m\geq0$}, the real components of the basis functions are shown whilst for \mbox{$m<0$} the imaginary components are shown. The solid purple (thick) lines mark the coefficients used by the MV pipeline to estimate the shear and flexions for an analysis with \mbox{$n_\mathrm{max}=10$}. The dashed purple (thick) lines mark the coefficients not used by the \citetalias{kui06} implementation for the same \mbox{$n_\mathrm{max}$}.}
\label{fig:polars}
\end{figure}

We introduce here an implementation of the Shapelets method which builds on a previous implementation described in \citet{kui06} (hereafter \citetalias{kui06}). This approach creates a Shapelets representation of the brightness profile of a PSF-convolved galaxy image. It also creates a model circular source and applies shear and flexion to it before convolving it with the point-spread function (PSF). Finally it fits the galaxy image to this modeled source in order to find the amount by which it has been sheared and flexed.

To first order in ellipticity $s$ and flexions $f$ and $g$, the model object can be written as
\begin{equation}
\mathbf{P}\cdot\left[1+\sum_{i=1,2}\left(t_i\hat{T}^i+s_i\hat{S}^i+f_i\hat{F}^i+g_i\hat{G}^i\right)\right]\sum_{\mathrm{even}}^{N_c}c_nC^n
\end{equation}
where $\mathbf{P}$ is the PSF matrix, $\hat{T}^i$, $\hat{S}^i$, $\hat{F}^i$ and $\hat{G}^i$ are the translation, shear, F flexion and G flexion operators respectively and $t_i$, $s_i$, $f_i$ and $g_i$ are the corresponding coefficients. The translation terms here ensure that fits spoiled by undue centroid shifts are caught. The operators are acting on a circular source which can be expressed as a series of circular Shapelets $C^n$ with coefficients $c_n$ where $n$ is even and the series is truncated at $N_c=n_{\mathrm{max}}-2$. The reason for truncating at $N_c$ rather than $n_{\mathrm{max}}$ is to safeguard against PSF structure at higher orders affecting the highest order Shapelets used. {To avoid introducing signal-to-noise (S/N) dependent biases, the} $n_{\mathrm{max}}$ is kept constant for all galaxies rather than being allowed to vary according to size or brightness. For faint sources, this means the higher-order coefficients will be noisy but unbiased.

Once we have a cartesian Shapelets representation of both the sheared, flexed and PSF convolved circular model and of the PSF convolved object we want to fit, we convert them both into polar Shapelets as described in \citetalias{kui06}. Polar Shapelets are simply cartesian Shapelets of order $n=a+b$ expressed in polar coordinates, resulting in polar Shapelets of order $n$ with angular order $m \leq n$ and $n+m$ even. The construction of these is discussed in \citet{ref03a} and further investigated in \citet{mas05} and \citet{mrr07}. In our implementation, the purpose of converting the model and object Shapelets expansions into polar Shapelets is to avoid truncation effects. \mbox{F flexion}, shear and \mbox{G flexion} operators acting on a polar Shapelet of order $(n,m)$ generate terms at order $(n\pm1,m\pm1)$, $(n\pm2,m\pm2)$ and $(n\pm3,m\pm3)$ respectively. By truncating the polar Shapelets expansion in the diamond shape shown in Figure~\ref{fig:polars}, i.e.~only including terms up to order $(N_c,0)$, $(N_c-1,\pm1)$, $(N_c-2,\pm2)$ and $(N_c-3,\pm3)$ in the fit, we minimise truncation effects from the mixing of orders.

The model is fit to each source using least-squares, resulting in a simultaneous estimate for the ellipticity $(s_1,s_2)$, the F flexion $(f_1,f_2)$, and the G flexion $(g_1,g_2)$. As explained in \citetalias{kui06}, the errors on the Shapelet coefficients are derived from the photon noise and propagated through the $\chi^2$ function for this fit. By differentiating the $\chi^2$ at the best-fit, we obtain the covariances between the fit parameters, resulting in proper error estimates.

In essence, the main development since \citetalias{kui06} is the addition of flexion to the model and the inclusion of higher order polar Shapelets ($m=\pm3$) in the fit.

\section{Testing the Pipeline}\label{sec:tests}
Several aspects of the pipeline, such as the choice of scale radius $\beta$, the method of PSF correction and the effect of noise on ellipticity estimates, have been thoroughly tested in \citetalias{kui06} as part of the development of the \citetalias{kui06} pipeline. In this section we will therefore focus on testing the recovery of shear and flexion.

\subsection{GREAT08}\label{sec:tests_great08}
As participants in the GRavitational lEnsing Accuracy Testing 2008 (GREAT08) challenge \citep{bsa09,bbb10}, we were able to contrast the shear measurement capability of the \citetalias{kui06} pipeline with that of the MV pipeline under different observing conditions. The challenge provided a large number of simulated sheared and pixelated galaxy images with added noise. The performance of the different shape measurement pipelines taking part was quoted in terms of a quality factor, or Q-value, defined as
\begin{equation}
Q = \frac{k_Q\sigma^2}{\langle(\langle \gamma_{ij}^m-\gamma_{ij}^t\rangle_{j\in k})^2\rangle_{ikl}}
\end{equation}
where $\sigma^2=\sigma_\mathrm{stat}^2+\sigma_\mathrm{syst}^2$ is a combination of the statistical spread in the simulations and the expected systematic errors. The superscripts $m$ and $t$ denote measured and true (input) values respectively and $\gamma_{ij}$ is the shear component $i$ for simulation image $j$. The differences between the measured and true shears are averaged over different input shear sets $k$ and simulation branches $l$. The whole expression is normalised by $k_Q$ so that a method with a purely statistical spread in the measured shears will have a Q-value of $k_Q$ which is the level desirable for future surveys. In the case of GREAT08, $k_Q=1000$ and $\sigma^2=10^{-7}$, giving a Q-value nominator of $10^{-4}$. With this definition current methods, like those that took part in the earlier Shear TEsting Programme (STEP) \citep{hvb06,mhb07}, generally achieve $10 \lesssim Q\lesssim 100$. This is sufficient for current weak lensing surveys. For a more in-depth discussion on the Q-value and its relation to the STEP parameters $m$ (multiplicative bias) and $c$ (additive bias), we refer to \citet{kmh08}.

The overall Q-value was similar for the \citetalias{kui06} and the MV pipelines, both in the LowNoise\_Blind competition ($Q\sim20$) and in the RealNoise\_Blind ($Q\sim25$). When broken down into the separate observing condition branches some differences became apparent. In general the MV pipeline did exceptionally well under ``good'' observing conditions, e.g.~for the high S/N branch or for well resolved galaxies. Our own simulations described in the next section will further test the dependence of the MV performance on different observing conditions.

\subsection{FLASHES}\label{sec:tests_flashes}
\begin{table}
\caption{The different branches of FLASHES. Four parameters are varied between the branches according to this table.}
\label{tab:branches}
\begin{tabular}{@{}lcccc}
\hline
 					& Intrinsic shape 	& Galaxy profile 	& S/N & PSF \\
\hline
Fiducial			& Round				& Gaussian			& 100	& Round		\\
Shape branch		& Elliptical		& Gaussian			& 100	& Round		\\
Profile branch 1	& Round				& Exponential		& 100	& Round		\\
Profile branch 2	& Round				& de Vaucouleur		& 100	& Round		\\
S/N branch 1		& Round				& Gaussian			& 8		& Round		\\
S/N branch 2		& Round				& Gaussian			& 20	& Round		\\
S/N branch 3		& Round				& Gaussian			& 40	& Round		\\
PSF branch			& Round				& Gaussian			& 100	& Elliptical\\
\hline
\end{tabular}
\end{table}

As there is no flexion simulation set publicly available to date, we create our own FLexion And SHEar Simulations (FLASHES). FLASHES are very similar to the GREAT08 simulations in several respects. First, each galaxy is generated on a grid, ensuring that there is no overlap of objects, thus avoiding deblending issues. Second, each simulation image consists of $10000$ such objects. Third, each galaxy is generated through the following sequence: (i) simulate a sheared and/or flexed (elliptical) galaxy model (depending on simulation branch); (ii) convolve with the PSF; (iii) apply the noise model. Four parameters are varied between the different FLASHES branches; the intrinsic galaxy shape, the light profile of the galaxies, the S/N of the galaxies and the shape of the PSF. These parameters are detailed below and summarised in Table~\ref{tab:branches}.

\subsubsection{Simulation Details}\label{sec:tests_flashes_details}
All parameters except for the intrinsic ellipticities are kept constant in each simulation image, and all images are created using Monte-Carlo selection. This is very similar to the process described in \citetalias{kui06} and in \citet{bbb10}, but with the photon trajectories being influenced by flexion as well as by shear if required.

The galaxies are modeled with S{\'e}rsic intensity profiles \mbox{$I_\mathrm{gal}\propto e^{-kr^{1/n}}$} \citep{ser68} with varying indices $n$. A S{\'e}rsic index of $n=0.5$ is a Gaussian profile whilst $n=1$ and $n=4$ are exponential and de Vaucouleur profiles respectively. Half of the FLASHES branches have intrinsically round galaxies whilst the other half consists of galaxies with intrinsic ellipticities picked randomly from the ellipticity distribution of objects in the COSMOS survey. There is no intrinsic flexion included. The PSFs applied to the simulations are nearly Gaussian with a Moffat profile $I_\mathrm{PSF}=(1+r^2/a^2)^{-m}$ of index $m=9$. In half of the branches, the PSF is round whilst in the other half it is elliptical in the horizontal direction with $e_{1,\mathrm{PSF}}=0.02$. To mimic the properties of the COSMOS survey, we use a PSF FWHM of 2.1 pixels and a PSF convolved galaxy size of 5.8 pixels which is the typical size of the galaxies we use in our COSMOS analysis. Finally there are four S/N branches, with S/N being defined as \mbox{Flux/(Flux error)}. It is expected that shape measurements will be less accurate at low S/N. For this reason the MV pipeline applies a S/N cut at 10 in general. The low S/N branch of 8 is designed to test how biased measurements are below this cut. The high S/N branch of 100 tests biases under near-perfect noise conditions.

The strength of the different distortions is picked randomly but with the following maximum values: $|\gamma_{1,2}| \leq 0.05$, $|\mathcal{F}_{1,2}| \leq 0.008 \; \mathrm{pixel}^{-1}$ and $|\mathcal{G}_{1,2}| \leq 0.02 \; \mathrm{pixel}^{-1}$. The value of each distortion component is kept constant across each image, but differs between the 30 images in each set, and between different sets.

\subsubsection{Simulation Results}\label{sec:tests_flashes_results}
\begin{figure}
\includegraphics[width=84mm,height=84mm,angle=270]{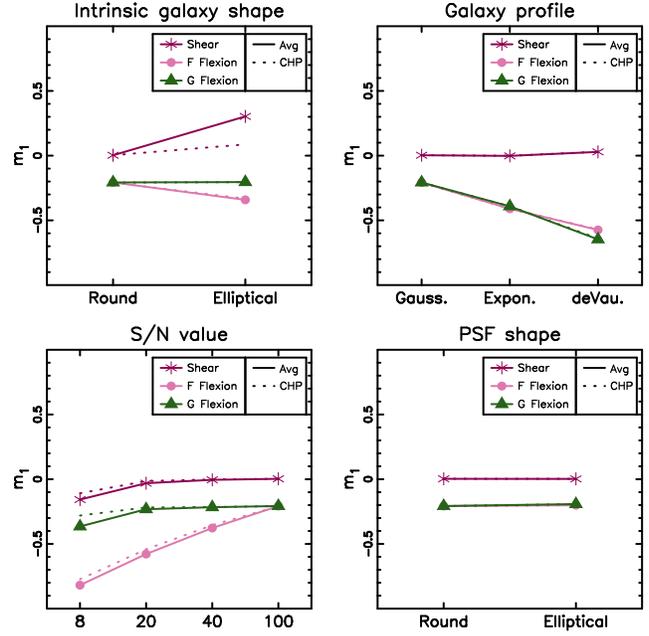}
\caption{ The multiplicative bias on the first component for each of shear, F flexion and G flexion. The purple stars represent shear, pink circles represent F flexion and green triangles represent G flexion. The symbols and solid lines show the weighted averages whilst the dashed lines show the CHP average. This is from running the MV pipeline on FLASHES, with $n_\mathrm{max}=10$. For the results for $m_2$, $c_1$ and $c_2$ please refer to Appendix~\ref{app:flashes}.}
\label{fig:flashes_results}
\end{figure}

To estimate the average distortion on each image we use two different techniques: a  weighted average with weights inversely proportional to the measurement errors, and Convex Hull Peeling (CHP). CHP is an efficient way of eliminating outliers and is essentially a 2D median. A convex hull, in the context of a point cloud in e.g.~the $\gamma_1,\gamma_2$ plane, is the minimal convex set of points containing that point cloud. Thus if all the points in this convex set were connected, a polygon containing the entire point cloud would be produced. By peeling away convex hulls, outliers are removed from the point cloud and the remaining points may be averaged over to produce a mean unaffected by extreme results. This is the averaging technique we used in GREAT08 where we peeled away 50\% of the measurements before averaging.

We employ the parameters $m$ and $c$ as used in STEP \citep{hvb06,mhb07} to quantify the performance of the software:
\begin{equation}
\langle\gamma^\mathrm{measured}_i\rangle-\gamma^\mathrm{input}_i = m_i\gamma^\mathrm{input}_i+c_i
\end{equation}\label{eq:step_mc}
and similarly for the flexions, where $i=1,2$ represents the shear component. A negative multiplicative bias $m_i$ thus indicates that the distortion is generally underestimated. A systematic offset $c_i$ may be caused by e.g.~insufficient PSF correction.

In Figure~\ref{fig:flashes_results} we show the multiplicative bias of the first component for each of shear, F flexion and G flexion as a function of the different simulation branches (please refer to Appendix~\ref{app:flashes} for the remaining bias components). For these results we use a Shapelets order of $n_\mathrm{max}=10$. We use {\tt SExtractor} \citep{ber96} to detect the objects in each simulation, which we then split into clean star and galaxy catalogues by matching to the input catalogue. We keep all properties apart from the one under investigation fixed at a fiducial value to allow for a fair comparison. The fiducial simulations in Figure~\ref{fig:flashes_results} have intrinsically round, high S/N galaxies with Gaussian light profiles and a circular PSF.

From the above figure it is clear that both flexions are likely to be underestimated, especially for higher S\'{e}rsic indices. The bias is also strongly S/N dependent, particularly for the F flexion. Thus a S/N cut is essential to improve the performance of the MV pipeline, but a bias correction should also be implemented. Investigating the dependence of $m$ on S/N further, we are able to fit the following power-law to our FLASHES results:
\begin{equation}
m_{1,2} = -a(\mathrm{S/N})^{-b}
\end{equation}
where $a$ and $b$ are constants as follows: for shear \mbox{$(a_{\gamma},b_{\gamma}) = (6.48,1.78)$}; for F flexion \mbox{$(a_{\mathcal{F}},b_{\mathcal{F}}) = (2.30,0.48)$}; for G flexion \mbox{$(a_{\mathcal{G}},b_{\mathcal{G}}) = (0.36,0.13)$}. We will apply this bias correction to our shape measurements in COSMOS, but since FLASHES have been tailored for this particular data the biases should be explored further before being applied to other surveys.

\subsection{Galaxy-galaxy Simulations and Bright Object Removal}\label{sec:tests_galgal}
\begin{figure}
\includegraphics[width=75mm,angle=270]{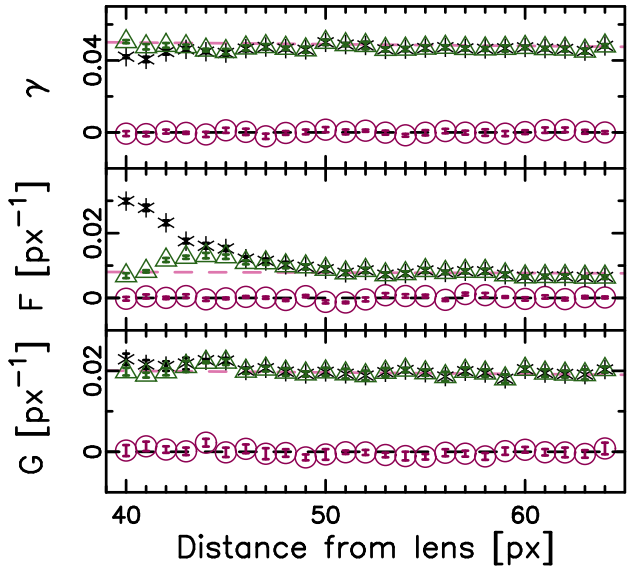}
\caption{ The shear (top panel), F flexion (middle panel) and G flexion (bottom panel) results from galaxy-galaxy lensing simulations, with and without Bright Object Removal (BOR). The black stars represent the tangential signal without BOR and the green triangles represent the same measurement corrected using BOR. The dashed pink line is the input signal and the purple circles are the cross-signal, which is expected to vanish, for the uncorrected measurements. Here, the FWHM of the lens is 14 pixels. Note the slight underestimation of the shear, the slight overestimation of the G flexion and the massive overestimation of the F flexion in the innermost bins when BOR is not applied.}
\label{fig:galgal_results}
\end{figure}

At the core of weak galaxy-galaxy lensing is the averaging of the signal in rings centered on lenses consisting of single galaxies rather than a galaxy cluster. This type of analysis is robust as numerous systematics, induced by e.g.~the PSF, cancel out. Different systematics may however be introduced, such as the light from the central, often bright, lens causing biases in the shape measurements as discussed in \citet{row08}. To study this possible effect, we created simple simulations with sources placed in evenly spaced rings around a central lens. Apart from source numbers and positions, the simulations were created in the same way as FLASHES. The S/N of the images was set to 200 to ensure minimum bias, and for the same reason the source galaxies had Gaussian light profiles. The size and profile parameters of the lens were varied between images.

The results for a lens with an exponential profile are shown in Figure~\ref{fig:galgal_results} (black stars), where we have used $n_\mathrm{max}=10$. We recover a near-perfect average signal in each source circle far from the lens. However, close to the lens the shear and G flexion are slightly affected, but, more strikingly, the F flexion is severely overestimated. The conclusion we draw from this is that {\em bright objects can add significantly to the F flexion signal}, due to light `leaking' into the Shapelets fitting radius. This causes the pipeline to detect a source light profile that is skewed towards the lens, and interpreting it as extra F flexion.

Our solution is to remove any bright objects sufficiently close to the source being fit using a technique we introduce here as Bright Object Removal (BOR). Before decomposing a galaxy image into Shapelets, we identify any bright objects that could conceivably intrude using selection criteria based on distance between the two objects, Shapelets fitting radius of the source, and size and brightness of the intruding object. We then create S\'{e}rsic models of the intruding objects using {\tt GALFIT} \citep{phi02} and subtract these models from the Shapelets stamp before doing the fitting. It works well in these simulations, provided one is careful with the parameters given to {\tt GALFIT} as input. The sky background value given to {\tt GALFIT} is particularly important as a small error in this estimate results in postage stamp artifacts when the stamps are subtracted from the original image. In Figure~\ref{fig:galgal_results} we also show the results if BOR is switched on whilst the rest of the analysis is kept identical to the previous run (green triangles). There is still some excess F flexion signal around 44 pixels, indicating that there may be some residual light remaining, but this excess is smaller than for the uncorrected measurements. This provides a confirmation that the measured signal reproduces the input signal well if BOR is applied, and no new artifacts are introduced. We note, however, that the leaking light does not affect the cross component of the measurements, with the consequence that this effect cannot be detected through the usual systematic checks.

\section{COSMOS Analysis}\label{sec:cosmos}
\citet{gol05} made a first detection of galaxy-galaxy flexion using the ground-based DLS, proving that flexion can indeed be detected, but ultimately they were hampered by the small size of their sample, the lack of redshifts and the extra blurring caused by the atmosphere. Therefore we choose the space-based Cosmic Evolution Survey (COSMOS, \citet{saa07}) as the first real dataset for the MV pipeline. Thanks to the depth of this survey we will have access to more than a thousand times as many lens-source pairs as \citet{gol05} did. More than half of these have photometric redshifts meaning that the division of the sample into lenses and sources will be more accurate. The intention is to provide independent confirmation that galaxy-galaxy flexion has high enough S/N to be detected, and that the software presented in this paper is able to do it. We will also  look closer to the lens than previous analyses and attempt to combine shear and flexion to give constraints on galaxy dark matter halo profiles.

\subsection{The COSMOS Data Set}\label{sec:cosmos_data}
COSMOS is to date the largest contiguous field imaged by the Hubble Space Telescope (HST) with a total area of 1.64~deg$^2$. The 579 tiles were observed in F814W (I-band) by the Advanced Camera for Surveys (ACS) between October 2003 and November 2005. Each tile consisted of 4 dithered exposures of 507 seconds each (2028 seconds in total) with about 95\% of the survey area benefiting from the full 4 exposures.

We use the images reduced by \citet{shj10} (hereafter \citetalias{shj10}) and also their catalogues for stars and galaxies, detected using {\tt SExtractor}. There are a total of 446~934 galaxies with $i_{814}<26.7$ in the mosaic catalogue, almost half of which have COSMOS-30 photometric redshifts from \citet{ics09}. These redshifts are magnitude limited and cover the entire COSMOS field to a depth of $i^+<25$.

\subsection{Data Analysis}\label{sec:cosmos_analysis}
Galaxy-galaxy lensing is less affected by the problems plaguing cosmic shear analyses, since most systematic shape distortions induced by instruments cancel out when azimuthally averaged. Still, we have to be careful not to introduce new systematic effects or biases, so correcting for the PSF and the charge-transfer inefficiency (CTI) \citep[e.g.][]{rma07,msl10} is important.

We use all galaxies with redshifts of $z<0.6$ as lenses. At higher redshifts the light from the lensing galaxies becomes difficult to account for due to the small angular separation on the sky, as explained further in Appendix~\ref{app:highz}. Furthermore, imposing a lens redshift cut will ensure that the vast majority of sources are truly background objects.

Our source catalogue is comprised of all objects with a shape measurement. We clean this catalogue using a series of conditions on size and measured shape, detailed in Appendix~\ref{app:cosmos_catalogues}, the most important of which is to remove objects with $\mathrm{S/N}<10$. Roughly two-thirds of the remaining sources have individual COSMOS-30 photometric redshifts assigned to them. For the remaining third (redshift bin 6 in \citetalias{shj10}) we use the estimated redshift distribution employed by \citetalias{shj10} to assign mean angular diameter distance ratios ($D_{s}/D_{ls}$) to each lens-source pair. We are finally left with $216\;873$ sources, corresponding to a source density of $\sim37 \; \mathrm{arcmin}^{-2}$. For the pairs we use, the median lens redshift is $z_{lens}=0.27$ and the median source redshift is $z_{source}=0.98$.

Despite the excellent space-based resolution, we need to correct the galaxy shapes for the instrumental PSF. The ACS PSF is known to fluctuate both spatially and temporally \citep[e.g.][]{rma07,ses07}, a variation mostly driven by changes in telescope focus caused for example by the breathing of the telescope. We can map the PSF using stars, but, in high-galactic latitude ACS fields typically only $\sim 10-20$ stars are present. This number is too low for the standard approach of a polynomial interpolation. Instead, we closely follow the analysis of \citetalias{shj10}, who conducted a principal component analysis (PCA) of the ACS PSF variation as measured in dense stellar fields. Details for the Shapelets implementation of PCA may be found in Appendix~\ref{app:cosmos_psf}.

A challenge with using CCD detectors in space is that they are not protected by the atmosphere. Exposed, they continuously get bombarded by radiation, causing deterioration of the chip surface. The imperfections created in this way act as charge traps which causes inefficiency in the moving of electrons to read-out. This effect is known as CTI \citep[e.g.][]{rma07,msl10}. As the electrons get trapped and then released at a later point, charge trails following objects are created in the read-out direction, effectively causing a spurious shear signal in that direction. Our correction for CTI again closely follows \citetalias{shj10}, who derive parametric corrections for the change in polarization for both galaxies and stars. For more details on this correction, please refer to Appendix~\ref{app:cosmos_cti}.

Once corrected, the galaxy-galaxy shear and flexion signals are weighted according to the geometric lensing efficiency of each lens-source pair. In the case of flexion there is an extra scale dependence of the signal. For the Navarro-Frenk-White (NFW) profile \citep*{nfw96}, the strength of the shear signal scales as
\begin{equation}
\gamma_{\mathrm{NFW}} \propto \frac{D_{l}D_{ls}}{D_{s}}
\end{equation}\label{eq:nfw_shear}
where $D_{l}$, ${D_{s}}$ and $D_{ls}$ are the angular diameter distances to the lens, to the source, and between lens and source respectively \citep{wri00}. The flexion signals scale as
\begin{equation}
\mathcal{F}_{\mathrm{NFW}},\mathcal{G}_{\mathrm{NFW}} \propto  \frac{D^2_{l}D_{ls}}{D_{s}}
\end{equation}\label{eq:nfw_flexion}
\citepalias{bgr06}. We therefore weight the signals accordingly, scale them to a reference lens and source redshift and compute the weighted average in 25 logarithmic distance bins (see Appendix~\ref{app:cosmos_signal} for details). We use a reference lens redshift of $z_{l,\mathrm{ref}}=0.27$ since that is close to the effective median redshift of our lenses, and a reference source redshift $z_{s,\mathrm{ref}}=0.98$. To estimate the errors on each bin and the covariances between them, we use 5000 bootstrap resamples of our source catalogue.

\subsection{Results}\label{sec:cosmos_results}
\begin{figure*}
\includegraphics[width=54mm,angle=270]{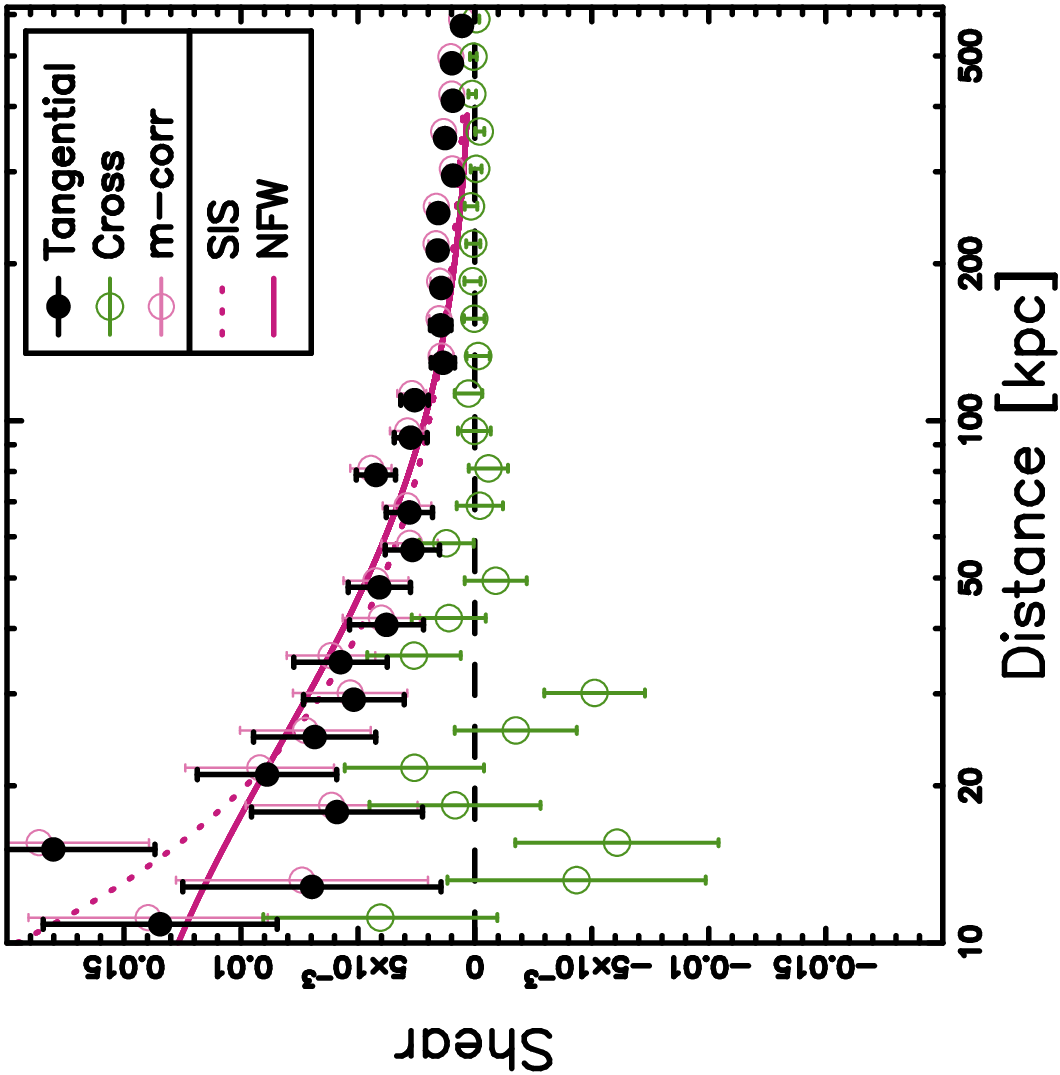}
\includegraphics[width=54mm,angle=270]{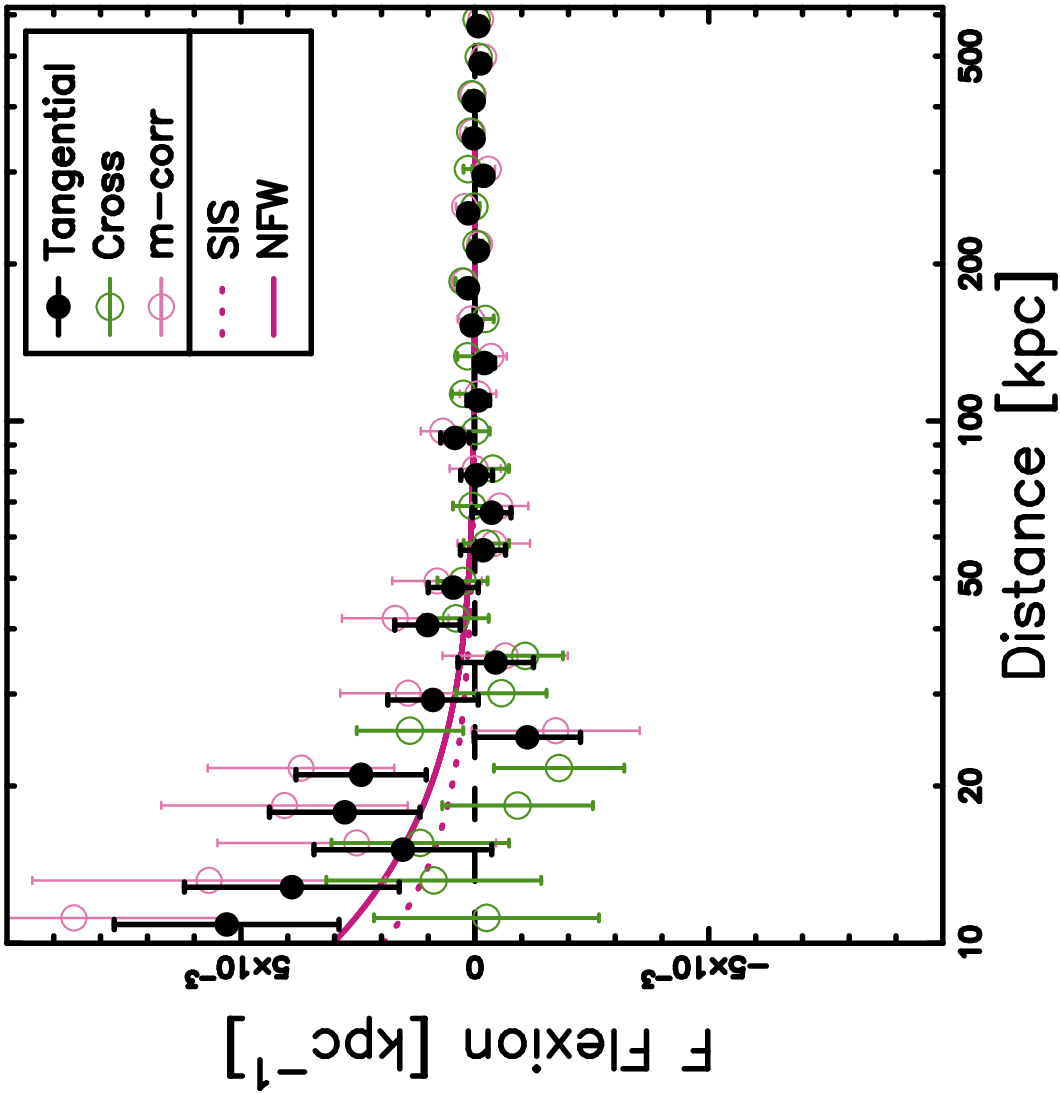}
\includegraphics[width=54mm,angle=270]{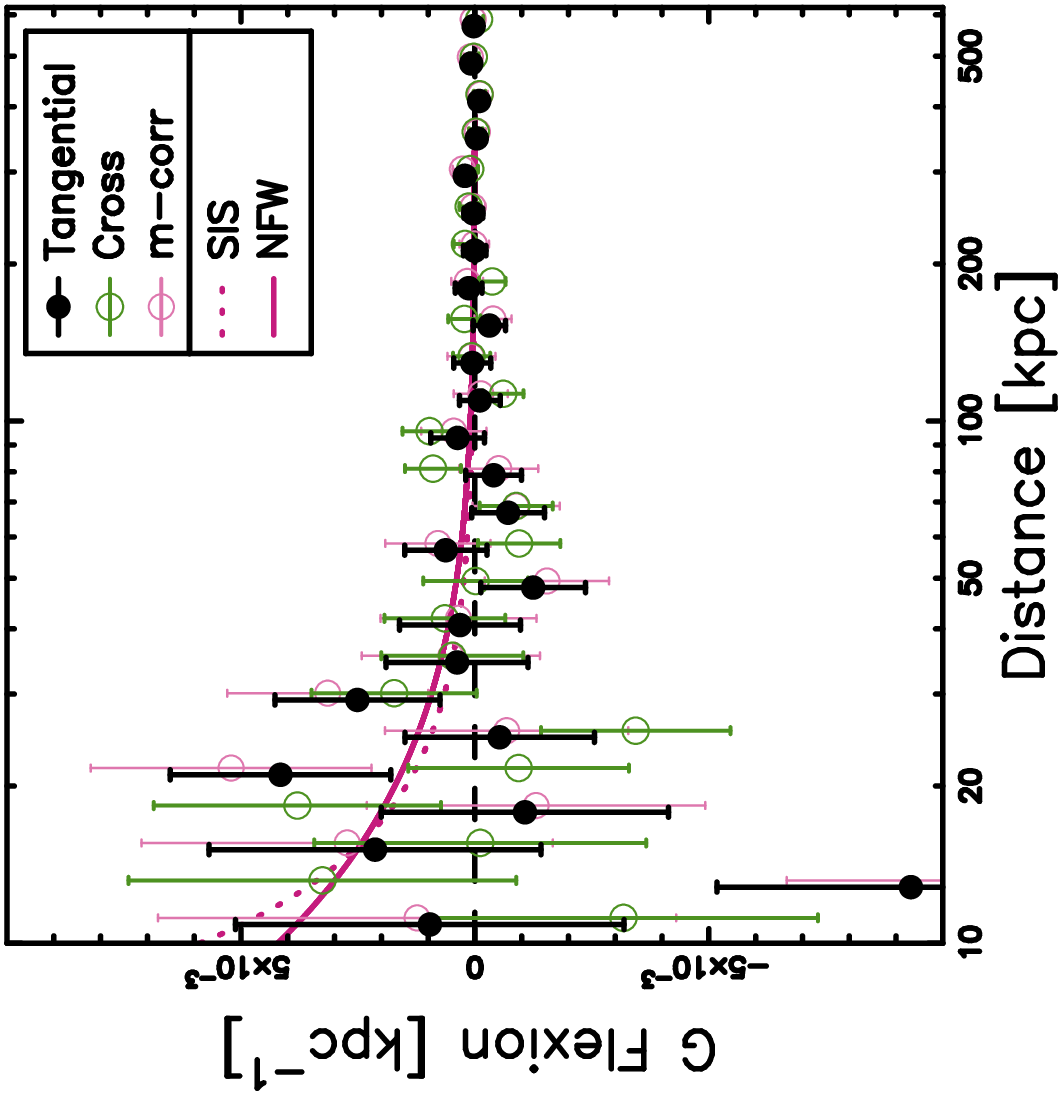}
\caption{ The galaxy-galaxy lensing results for the COSMOS data, using a maximum Shapelets order of \mbox{$n_\mathrm{max}=10$}. Black solid points represent the tangential signal and green circles represent the cross term. The pink circles represent the tangential signal if we apply the multiplicative bias correction implied by FLASHES. Note that the SIS and NFW profiles have been fitted to the shear data and then {\em translated} into predictions for $\mathcal{F}$ and $\mathcal{G}$ curves.}
\label{fig:cosmos_results}
\end{figure*}

The results from our galaxy-galaxy lensing analysis of the full COSMOS lens and source sample is shown in Figure~\ref{fig:cosmos_results}. In the left panel we plot the shear results as a function of physical distance from the lens. These results agree very well with those from \citetalias{shj10} (see Appendix~\ref{app:ksb}), providing an independent consistency check. To this we fit a Singular Isothermal Sphere (SIS) profile and a tentative NFW profile. Due to the dependence on mass and redshift of the mean concentration parameter \citep[e.g.][]{dsk08}, the NFW profile is only an indication when the spread in lens masses and redshifts is as great as it is in the above sample. Splitting the sample up into redshift and/or mass bins would increase the confidence in the fit, but decrease the S/N of the signals significantly.

The middle and right panels show the F and G flexion results respectively, for the same lenses and sources. The profiles plotted here are identical to those plotted in the shear panel but {\em translated} into predictions for $\mathcal{F}$ and $\mathcal{G}$, as opposed to fitted to the flexion data directly. The F flexion has a tendency to be overestimated compared to the predicted profile from the shear, and we investigate this discrepancy further in the following sections. We also note that we measure a G flexion that is very noisy and consistent with zero. This is most likely caused by lack of information in higher $m$-order Shapelets for fainter sources, and we choose to use only shear and F flexion in the continued analysis.

Also shown in pink circles in Figure~\ref{fig:cosmos_results} is the signal if we apply the multiplicative S/N-dependent bias correction implied by FLASHES. With this correction, the F flexion signal becomes slightly higher. This bias correction is only based on one specific set of simulations and is thus rather preliminary; this is also indicated in the increased size of the error bars. Correcting for the morphology-dependent bias requires accurate source morphology determination. Using the photometric galaxy type estimates from \citet{ics09} as an indicator of morphology we find that $<5\%$ of our source sample consists of likely de Vaucouleur objects. This type estimate is not accurate enough to implement a morphology bias correction, but simply removing the de Vaucouleur candidates we identified makes little difference to our results. It is clear, however, that an accurate bias calibration of the flexion amplitude, taking into account both source S/N and brightness profiles, requires further investigation.

\subsection{Removing Bright Objects}\label{sec:cosmos_removal}
\begin{figure}
\includegraphics[width=60mm,angle=270]{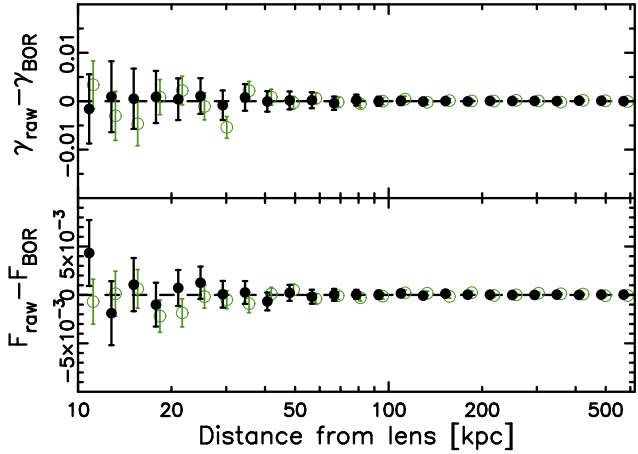}
\caption{ Comparison between the galaxy-galaxy shear and flexion signals with and without Bright Object Removal, showing the non-zero correction to the innermost F flexion bin (corresponding to roughly \mbox{40 px} in Figure~\ref{fig:galgal_results}). Black solid points represent the difference between the signals before and after correction, with the F flexion in units of $\mathrm{kpc}^{-1}$, whilst green circles represent the cross term.}
\label{fig:cosmos_brightbin}
\end{figure}

We now explore the tendency of the F flexion points to lie above the predicted profiles. As shown in Section~\ref{sec:tests_galgal}, the shape measured may be affected by bright objects nearby. We implement BOR in our COSMOS analysis to see the effect on real data. For very well resolved objects, prominent spiral arms and other complications cause {\tt GALFIT} to reject the single S\'ersic profile fit. Removing these objects, and the residual light from the wings of the profile (Figure~\ref{fig:galgal_results}), requires a more sophisticated model. For now we are only interested in a rough indication of the impact this light leakage has on a galaxy-galaxy signal so we will not correct for the few large objects in this paper. However, as shown in Figure~\ref{fig:cosmos_brightbin}, the correction to the innermost F flexion bin is non-zero even without accounting for the very large objects. The shear is largely unaffected, but for flexion analyses in future deeper and larger surveys it will be important to correct for this effect.

\subsection{The Effect of Substructure}\label{sec:cosmos_substructure}
\begin{figure}
\includegraphics[width=75mm,angle=270]{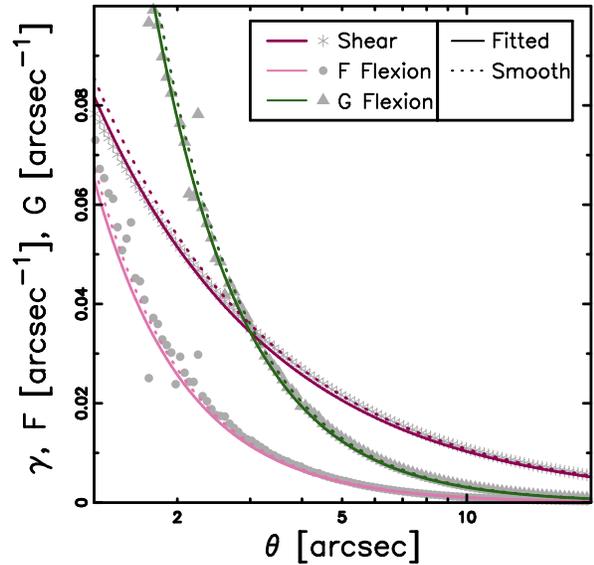}
\caption{ Simulated shear and flexion signals azimuthally averaged in galaxy haloes with and without TSIS subhaloes. Grey stars, circles and triangles represent the binned shear, F flexion and G flexion respectively. Purple, pink and green lines represent the shear, F flexion and G flexion signal if the halo is a smooth SIS (dashed). The solid lines are an SIS profile as fitted to the shear data points in a simulated galaxy containing TSIS subhaloes and translated into predictions for the flexions.}
\label{fig:tsis}
\end{figure}
Since flexion is more sensitive to the underlying mass distribution on small scales than shear is, we expect it to respond differently to the presence of substructure in galaxy haloes. To test whether this has any impact on our analysis we take a galaxy-size SIS halo \citepalias[see][for shear and flexion expressions]{bgr06} and populate it with subhaloes, allowing 20\% of the mass to be in substructure. The total mass of the halo is $10^{12}\;h^{-1}M_{\odot}$ and the galaxy is placed at $z=0.35$ with $D_l/D_{ls}=0.5$. We spread the substructure mass over 100 subhaloes, randomly distributed according to an SIS density profile. Finally we average the azimuthally averaged signal over 100 such galaxies. Now, subhaloes are generally stripped. To approximate this we use a Truncated SIS (TSIS) profile for the subhaloes \citep[see][for constraints on parameters]{hyg04}. The TSIS convergence is given by 
\begin{equation}
\kappa(\theta) = \frac{\theta_E}{2\theta}\left(1-\frac{\theta}{\sqrt{\theta^2+\theta_S^2}}\right)
\end{equation}
where $\theta_E$ is the Einstein radius and $\theta_S$ is a truncation scale where the profile steepens. On small scales ($\theta \ll \theta_S$) the TSIS behaves like an SIS but at large scales ($\theta \gg \theta_S$) the profile decreases as $\theta^4$. The TSIS shear is given in \citet{sch97} and the flexions are
\begin{equation}
\mathcal{F}(\theta) = \frac{\theta_E}{2\theta^2}\left(\frac{\theta^3}{(\theta^2+\theta_S^2)^{3/2}}-1\right)e^{i\phi}
\end{equation}
and
\begin{equation}
\mathcal{G}(\theta) = \frac{\theta_E}{2\theta^3}\left(3\theta+8\theta_S-\frac{3\theta^4+12\theta^2\theta_S^2+8\theta_S^4}{(\theta^2+\theta_S^2)^{3/2}}\right)e^{i3\phi}
\end{equation}
where $\phi$ is the position angle of the background source. Using the parameters above and a truncation scale \mbox{$\theta_S=2\;\mathrm{arcsec}$} for the subhaloes we get the results shown in Figure~\ref{fig:tsis}. The shear profile fit is pulled down slightly compared to a smooth halo but the flexions are not similarly affected. Due to the substructure the flexions are more scattered, but the overall trend is for the points to follow the smooth profile, or even slightly above in the F flexion case. Thus the flexions seem overestimated compared to the shear fit. We stress however that the fraction of substructure used in this test (20\%) is high to exaggerate the effect. The test does show that substructure may affect the flexions differently to the shear, but its influence is likely less than the excess currently observed in COSMOS.

\subsection{Profile Determination}\label{sec:cosmos_profile}
\begin{figure}
\includegraphics[width=75mm,angle=270]{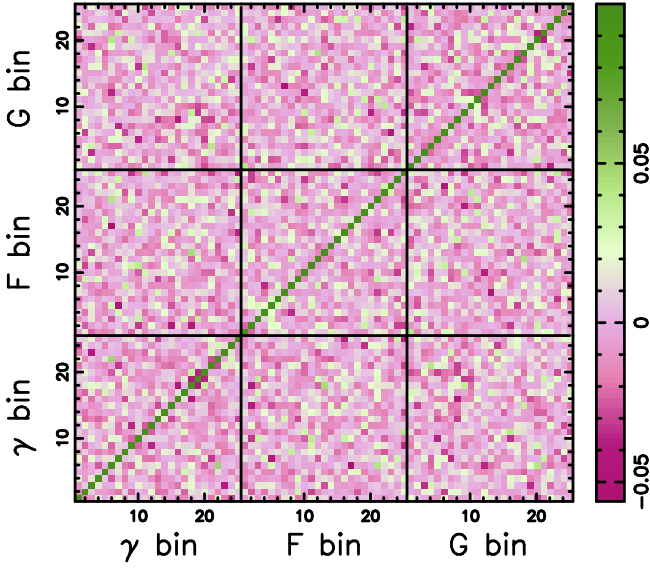}
\caption{ The correlation matrix between the shear and flexion bins, using 5000 bootstrap resamples. Please note the scale; to display the minute variations between off-diagonal elements we have artificially set {\em the diagonal elements (dark green) only} to 0.1, whilst all other elements are unscaled and normalised to diagonal elements of 1.0 as is customary.}
\label{fig:cosmos_correlation}
\end{figure}
\begin{figure}
\includegraphics[width=75mm,angle=270]{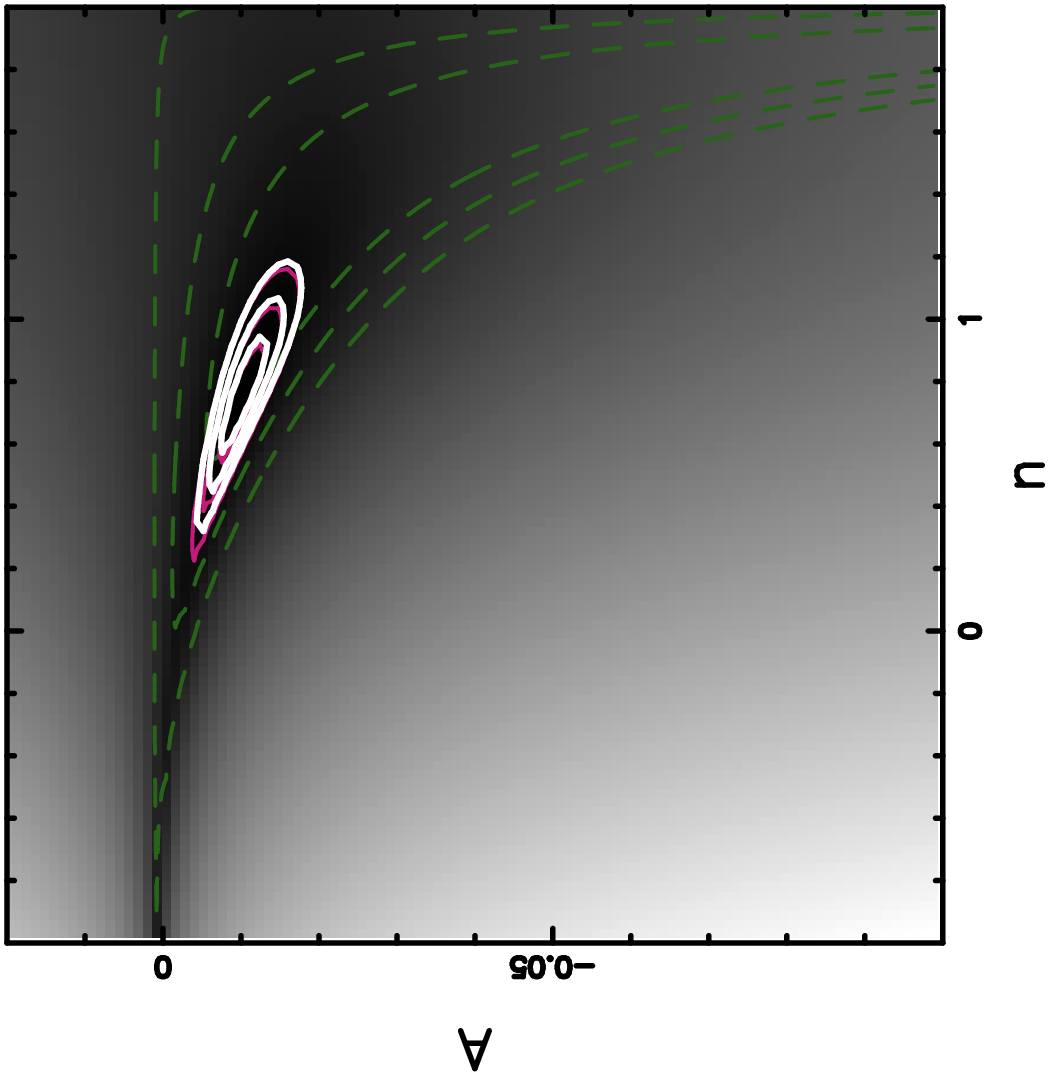}
\includegraphics[width=75mm,angle=270]{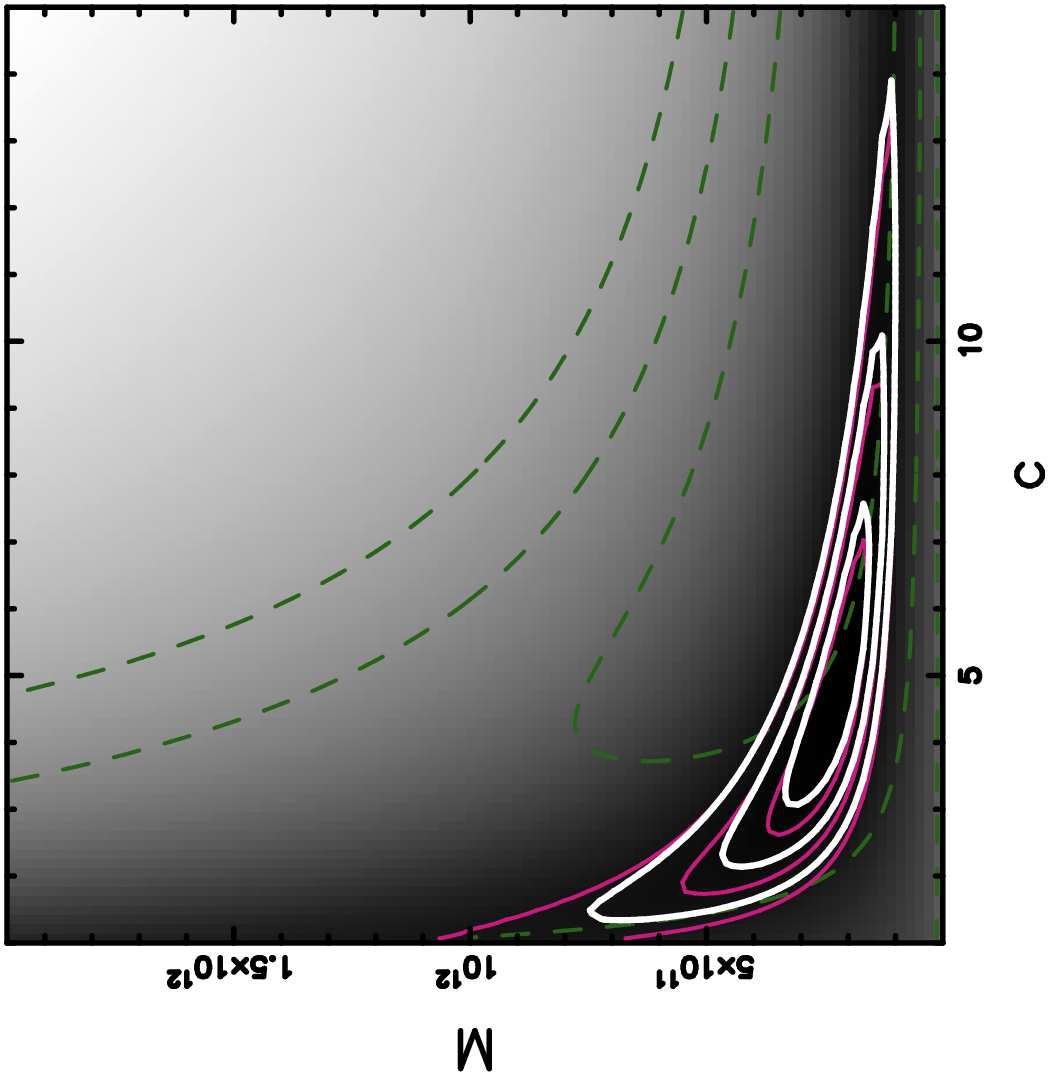}
\caption{ Joint profile constraints using shear and F flexion. The top (bottom) panel shows the results for the power-law (NFW) fit. Purple (thin solid) lines represent shear and green (dashed) represent F flexion. The contours show the 67.8\%, 95.4\% and 99.7\% confidence limits respectively in terms of constant $\Delta\chi^2$ ($2.30$, $6.17$ and $11.8$ respectively). The white (thick) contour marks the joint confidence limits. The grey-scale is logarithmic in $\chi^2$.}
\label{fig:cosmos_profiles}
\end{figure}

One of the most interesting potential uses of flexion is as an aid to shear in determining the inner density profiles of dark matter haloes. The two signals are sensitive to the underlying density profile on different scales, so combining the two will give us tighter constraints than either on their own. To combine the shear and flexion signals we have to take any correlation between them into account. \citetalias{bgr06} assumed that the shear and flexion measurements would be uncorrelated. Here we confirm this assumption through the correlation matrix between the shear and flexion bins, using 5000 bootstraps, shown in Figure~\ref{fig:cosmos_correlation}. This implies that it is trivial to combine the shear and flexion information to find the profile of an average lens. We use the F flexion in conjunction with the shear to fit density profiles to the measured signal. For this purpose we try two different families of profiles: the power-law and the NFW. Our general power-law is defined as
\begin{equation}
\gamma = -Ad^{-n}
\end{equation}\label{eq:powerlaw_shear}
with $d$ the distance from the lens, and the amplitude $A$ and the index $n$ free parameters. An index of $n=1$ would be equivalent to an SIS. The above expression is easily differentiated to give the F flexion
\begin{equation}
\mathcal{F} = (n-2)Ad^{-n-1}.
\end{equation}\label{eq:powerlaw_flexion}
The expressions for the NFW profiles are somewhat more complicated but they are given in full in \citet{wri00} and \citetalias{bgr06} for shear and flexion respectively. Here we leave the virial radius $M_{200}$ and the concentration $c$ as free and independent parameters. We fit the power-law and NFW profiles to the inner \mbox{100~kpc} only as this is the region where \mbox{F flexion} becomes important and the shear profile is not affected by halo-halo contamination.

The top panel in Figure~\ref{fig:cosmos_profiles} shows that both the shear and the \mbox{F flexion} are consistent with an SIS ($n=1$), although together they prefer a slightly lower power-law index of \mbox{$n=0.73^{+0.40}_{-0.43}$}. The bottom panel shows that it is difficult to constrain the NFW concentration if it is left completely unrestricted. This analysis with two free and independent parameters is not completely representative, however, since simulations indicate a fixed mean mass-concentration relationship \citep{dsk08}. It is also important to note that the average profile we constrain here is a composite of lenses in a large redshift range. Detection at the high end of the redshift distribution tend to be biased towards intrinsically brighter objects than at the low end. We also combine measurements from lenses of different sizes and morphologies. Nonetheless, combining shear and F flexion does provide tighter constraints than shear alone on the density profiles, and this is an important proof of concept. The resulting mass estimate for the average lens in COSMOS from the combined NFW fit is \mbox{$M_{200}=2.12^{+3.60}_{-1.09}\times10^{11}\;h^{-1}M_{\odot}$} with a concentration of \mbox{$c=4.82^{+7.04}_{-3.16}$}.

\section{Discussion and Conclusions}\label{sec:discussion}
We have shown a significant detection of galaxy-galaxy F flexion for the first time with Shapelets using the space-based COSMOS data set. We used this flexion signal in conjunction with the shear to constrain the average density profile of the galaxy haloes in our lens sample. We found a power-law profile consistent with an SIS. Furthermore, we showed that the inclusion of F flexion provides tighter constraints on both power-law and NFW profiles, an important proof of concept.

The galaxy-galaxy F flexion signal measured in COSMOS is slightly higher than expected from the shear signal, especially if we apply the multiplicative bias correction. There is however no indication from the cross term that there are systematics present. The discrepancy could be partly due to insufficient nearby object light removal, but this is unlikely to explain the full offset. Substructure in galaxy haloes may cause excess F flexion compared to what the shear measures. However, a large fraction of the galaxy halo mass has to be in substructure in order for the effect to become significant. We note that \citet{gol05} also find shear and F flexion signals that are inconsistent with each other; the velocity resulting from an SIS profile fit to their F flexion signal is nearly twice that found using shear. This is qualitatively consistent with our findings, which leads us to believe that there is something more fundamental affecting the signal. In the near future we would like to further investigate the dependence of these discrepancies on lens properties.

We measure a galaxy-galaxy G flexion signal that is consistent with the predicted profile, but due to the large measurement errors it is also consistent with zero. This measurement is a lot noisier than the other two, an effect most likely caused by the fact that there is less information available in the higher $m$-order Shapelets for fainter sources. To measure a G flexion signal we thus require many well-resolved sources, an extravagance not yet awarded us. Future large space-based surveys such as EUCLID will enable us to investigate G flexion further, but for now F flexion is a promising tool in its own right.

The software introduced in this paper, the MV pipeline, is able to detect these higher order lensing distortions. We have shown that in practice, the Shapelets F flexion measure is affected by light from nearby bright objects and detailed a way to correct for this effect. This BOR does require further sophistication to account for large, well resolved galaxies, galaxies which are not well described by the single S\'{e}rsic light profile employed here. From the FLASHES simulations it is clear that there is more work required in order to improve the accuracy of the F flexion measurements for future surveys. Noise related biases are particularly significant for this type of shape measure, and we have modeled these biases in COSMOS.

In the future we hope to measure flexion on a larger survey, enabling us to reduce the noise so that we can investigate the trend with e.g.~redshift and lens mass. A larger number of sources would also enable us to further tighten the profile constraints in the inner regions of dark matter haloes where baryons become important. It is not yet clear how well we can measure flexion on ground-based data, but surveys like KiDS, CFHTLS and RCS2 should provide an excellent test-bed.

\section*{Acknowledgments}
We would like to thank our colleagues Henk Hoekstra, Edo van Uitert and Elisabetta Semboloni at Leiden Observatory, and Peter Schneider at Bonn University, for useful discussions. Gary Bernstein drew our attention to the possibility of using Convex Hull Peeling for our averages, and for this we would like to thank him. MV is supported by the European DUEL Research-Training Network (MRTN-CT-2006-036133). TS acknowledges support from the Netherlands Organization for Scientific Research (NWO). 

\bibliographystyle{mn2e}

\newpage

\appendix

\section{FLASHES Results}\label{app:flashes}
\begin{figure}
\includegraphics[width=84mm,height=84mm,angle=270]{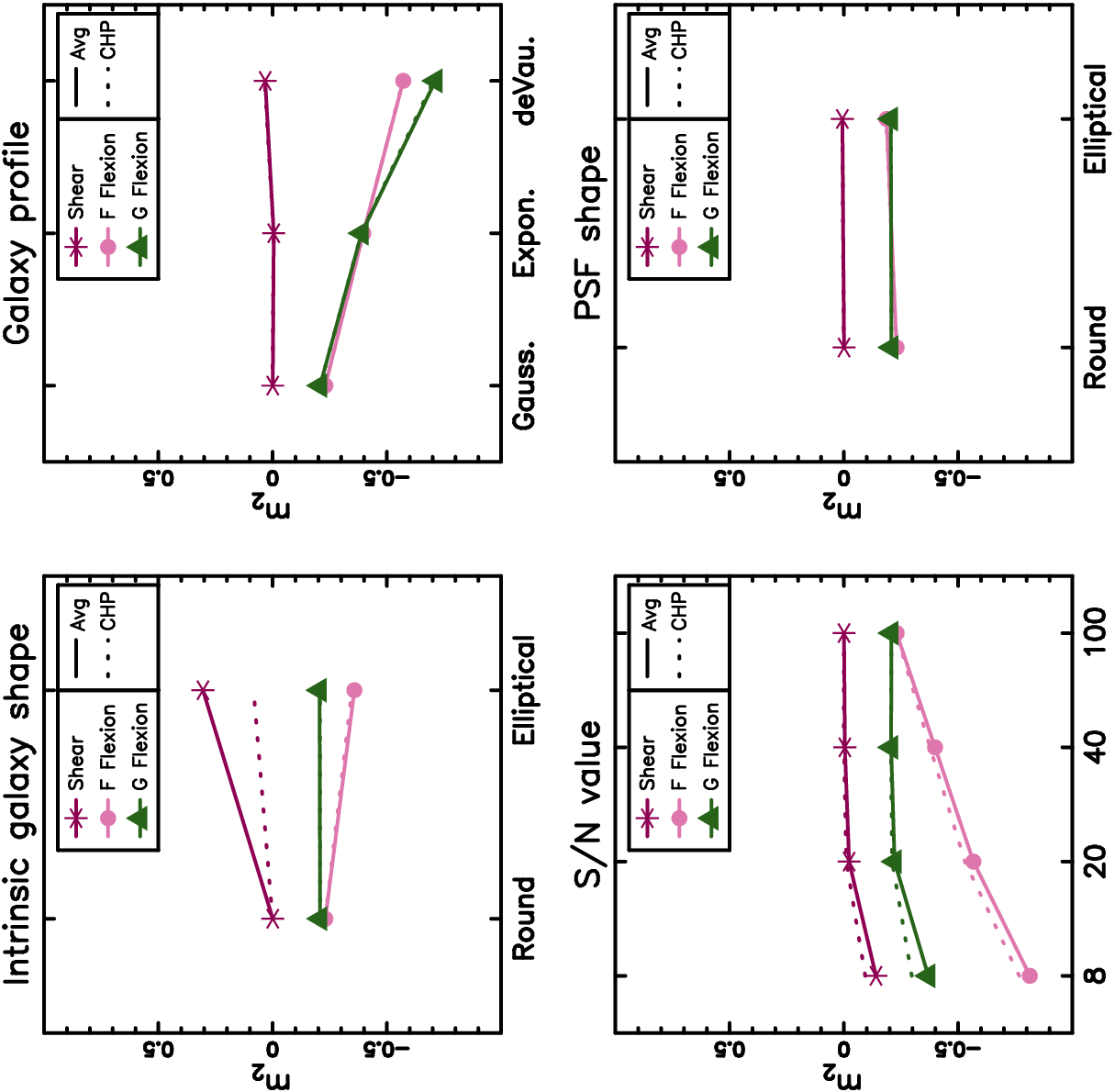}
\caption{ The multiplicative bias on the second component for each of shear, F flexion and G flexion. The purple stars represent shear, pink circles represent F flexion and green triangles represent G flexion. The symbols and solid lines show the weighted averages whilst the dashed lines show the CHP average.}
\label{app:fig:flashes_results_m2}
\end{figure}
\begin{figure}
\includegraphics[width=84mm,height=84mm,angle=270]{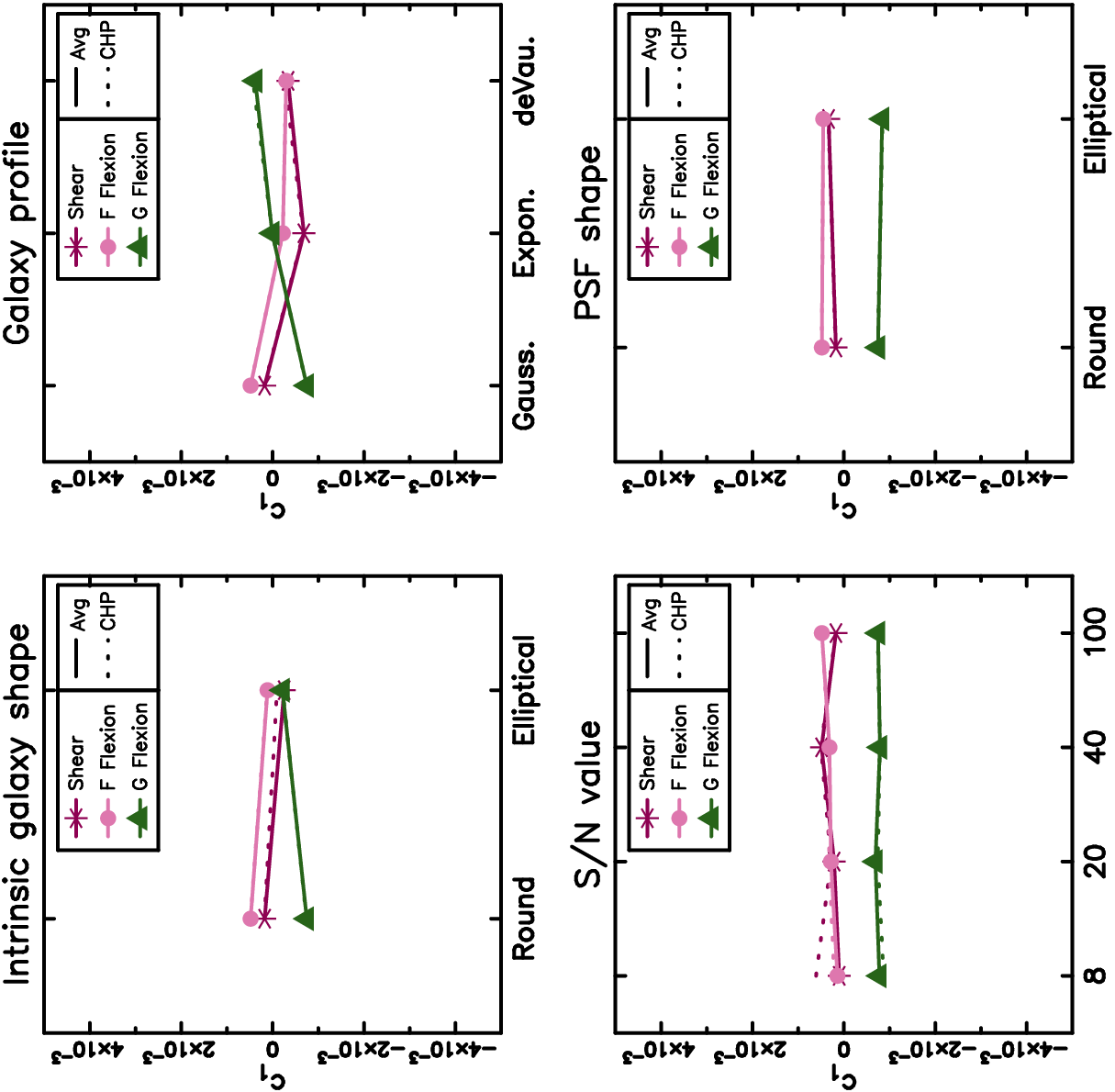}
\caption{ The additive bias on the first component for each of shear, F flexion and G flexion. As before, the purple stars represent shear, pink circles represent F flexion and green triangles represent G flexion. The symbols and solid lines show the weighted averages whilst the dashed lines show the CHP average.}
\label{app:fig:flashes_results_c1}
\end{figure}
\begin{figure}
\includegraphics[width=84mm,height=84mm,angle=270]{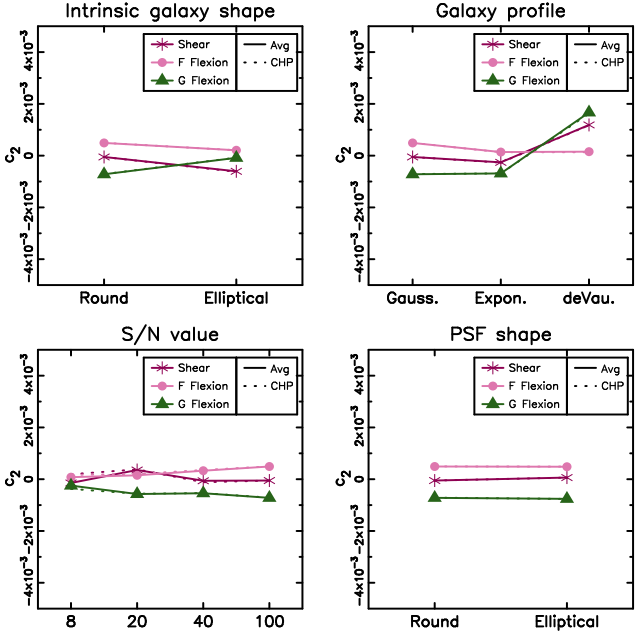}
\caption{ The additive bias on the second component for each of shear, F flexion and G flexion. As before, the purple stars represent shear, pink circles represent F flexion and green triangles represent G flexion. The symbols and solid lines show the weighted averages whilst the dashed lines show the CHP average.}
\label{app:fig:flashes_results_c2}
\end{figure}

The figures shown in this Appendix complement Figure~\ref{fig:flashes_results} in the main paper and provide additional detail on the results from running the MV pipeline on FLASHES, with $n_\mathrm{max}=10$. The parameters $m$ and $c$ are defined through
\begin{equation}
\langle\gamma^\mathrm{measured}_i\rangle-\gamma^\mathrm{input}_i = m_i\gamma^\mathrm{input}_i+c_i
\end{equation}
and similarly for the flexions, where $i=1,2$ is the component. We use two different techniques to estimate the average distortion on each image: a weighted average and Convex Hull Peeling (CHP).

In Figure~\ref{app:fig:flashes_results_m2} we show the multiplicative bias of the second component for each of shear, F flexion and G flexion as a function of the different simulation branches. For these results we use a Shapelets order of $n_\mathrm{max}=10$. This bias behaves as the multiplicative bias of the first component (Figure~\ref{fig:flashes_results}), as expected. The biases of all distortion measurements, and in particular F flexion, are severely dependent on S/N and brightness profile.

The additive bias $c$ is minimal for shear and F flexion (see Figures~\ref{app:fig:flashes_results_c1} and \ref{app:fig:flashes_results_c2}) indicating that the PSF is either well corrected for or not significantly affecting these two measurements. For the G Flexion the offset is larger.

\section{COSMOS Data Analysis}\label{app:cosmos_analysis}

\subsection{Catalogue Creation}\label{app:cosmos_catalogues}
To maximise the number of lens-source pairs we use all objects with assigned photometric redshifts as sources, but imposing a redshift cut of \mbox{$z<0.6$} for lenses. Additionally we use sources without individual redshifts (S10 redshift bin 6), assigning mean angular diameter distance ratios ($D_{s}/D_{ls}$) to these lens-source pairs according to the estimated redshift distribution employed by S10. We then weight all pairs with their individual lensing efficiency, similar to the weighting scheme in e.g.~\citet{msk06} (see Appendix~\ref{app:cosmos_signal}). This downweights pairs that are close in redshift and naturally removes pairs where the ``source'' is in front of the ``lens''. To the source catalogues we apply the following cuts:
\begin{itemize}
\item $S/N > 10$. This cut is important as the F flexion measurement in particular gets heavily biased towards low S/N (see Section~\ref{sec:tests_flashes_results}).
\item If the centroid cannot be determined accurately the Shapelets decomposition will be inferior. Therefore objects where the code is forced to move the centroid compared to the one estimated by {\tt SExtractor} by more than half a pixel are excluded.
\item The summed power in constant $m$ of the polar Shapelets provides an indicator of the Shapelet fit being affected by a neighbouring object. If the fractional power is particularly high at high orders the object is excluded \citepalias[see][for more details]{kui06}.
\item If the FWHM or scale radius of the object is too small compared to the scale radius of the PSF the object is excluded.
\item If $\gamma^2>1.4$, $\mathcal{F}^2>3.0\;\mathrm{arcsec}^{-1}$ or $\mathcal{G}^2>6.6\;\mathrm{arcsec}^{-1}$ then the object is excluded. These numbers are based on the measured distributions and the cuts are applied to remove outliers with very noisy shape measurements.
\item Finally, we remove faint objects with an assigned photometric redshift of $z<0.6$ that have a prominent secondary peak at $z_{2\mathrm{nd}}>0.6$, as discussed in \citetalias{shj10}.
\end{itemize}

\subsection{PSF Interpolation}\label{app:cosmos_psf}
The ACS PSF fluctuates both spatially and temporally \citep[e.g.][]{rma07,ses07}, a variation mostly driven by changes in telescope focus caused for example by the breathing of the telescope. We can map the PSF using stars, but in high-galactic latitude ACS fields typically only $\sim 10-20$ stars are present. This number is too low for the standard approach of a polynomial interpolation. Instead, we closely follow the analysis of \citetalias{shj10}, who conducted a principal component analysis (PCA) of the ACS PSF variation as measured in dense stellar fields. They found that $\sim 97\%$ of the PSF variation can be described with a single parameter (the first principal component). This parameter is related to the HST focus position, and we therefore dub it `focus'\footnote{The capturing of small additional variations beyond focus was relevant for the cosmic shear analysis of \citetalias{shj10}. Here we can safely ignore these minor additional effects. Galaxy-galaxy lensing is much less sensitive to PSF anisotropy residuals as they cancel out to first order.}

Here we make use of the \citetalias{shj10} measurement of the HST focus in all COSMOS exposures and the investigated stellar field exposures. We also obtain Shapelets versions of the focus-dependent \citetalias{shj10} PSF models, by decomposing the dense stellar field stars into Shapelets and interpolating between them with polynomials which are varied both spatially and with different powers of the focus principal component coefficient. From these models and from the COSMOS focus estimates we then compute a Shapelets PSF model for each COSMOS exposure, which we then combine to obtain a model for the stacked PSF at all galaxy positions.

\subsection{CTI Correction}\label{app:cosmos_cti}
\begin{figure}
\includegraphics[width=43mm,angle=270]{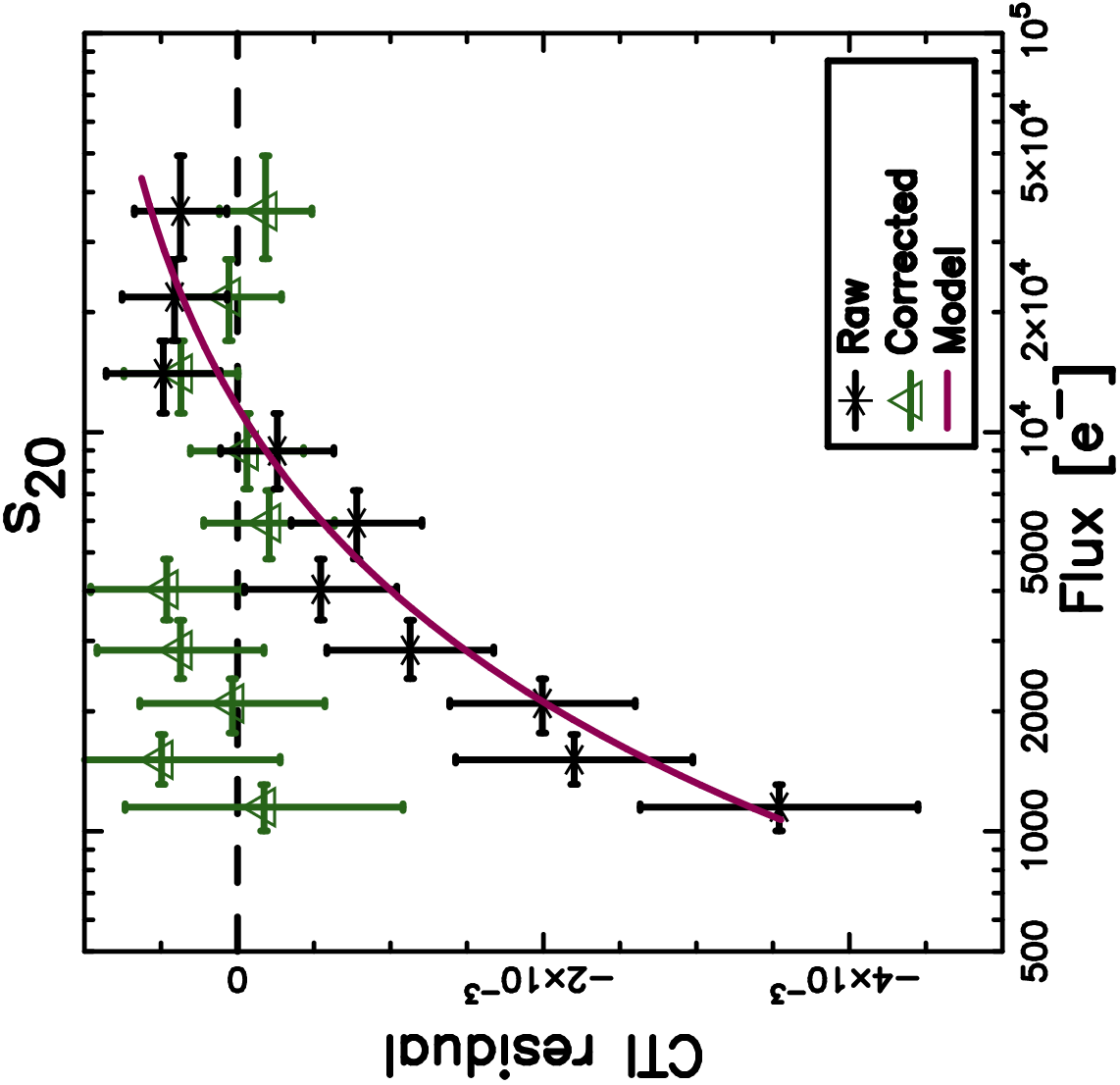}
\includegraphics[width=43mm,angle=270]{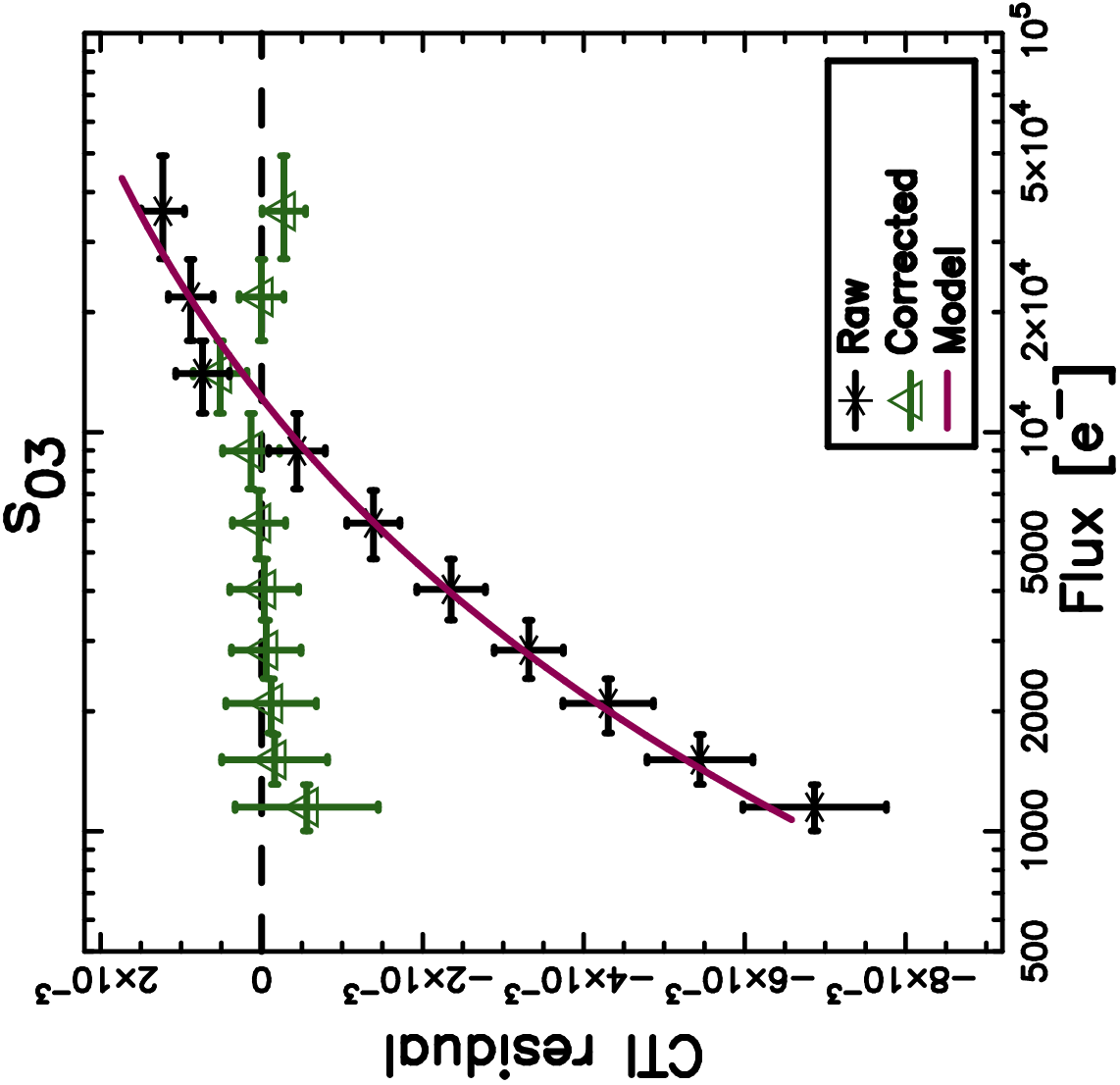}
\caption{CTI-induced residuals on the stellar Shapelet coefficients $s_{20}$ (left) and $s_{03}$ (right) in an example star field. The black stars show the mean of the coefficients as a function of stellar flux after subtraction of a spatial third-order polynomial model derived from bright stars to separate PSF and CTI effects. Each coefficient has been scaled to a reference number of $y_\mathrm{trans}=2048$ parallel readout transfers.
The purple curves show the parametric CTI model, jointly determined from 700 stellar field exposures. The horizontal dashed line indicates an offset corresponding to the mean CTI model for the bright stars used for the polynomial interpolation. The green triangles indicate the corrected coefficients after subtraction of the CTI model.}
\label{fig:cte}
\end{figure}

Our correction for CTI again closely follows \citetalias{shj10}, who derive parametric corrections for the change in polarization for both galaxies and stars. The correction for stars is important in order to measure the actual PSF, independent of the non-linear CTI effects. In the stellar field analysis we therefore correct the PSF cartesian Shapelet coefficients for CTI before generating the PCA PSF model. In order to estimate the influence of CTI on the different Shapelet coefficients, we follow \citetalias{shj10} and spatially fit each coefficient within one exposure with polynomials. Due to the limited depth of the charge traps, CTI is non-linear, and has a larger relative impact on faint sources than on bright ones. The CTI effect can thus be estimated from the flux-dependent residuals, after the polynomial model has been used to subtract both the flux-independent PSF and the flux-averaged CTI signal. 

Figure~\ref{fig:cte} shows these residuals as a function of stellar flux for the stellar Shapelets coefficients $s_{20}$ and  $s_{03}$ in one example stellar field. Here the residuals were scaled to the same number of readout transfers (2048). The CTI effect on the coefficients is clearly visible (black stars), but with our power law model (curve) it can be well corrected for (green triangles). The model is fit simultaneously from all 700 stellar fields as a function of stellar flux, sky background, time and number of readout transfers (see S10). CTI affects object shapes in the readout direction, which also after drizzling roughly matches the $y$-direction. Thus CTI residuals are expected to be roughly symmetric about the $y$-axis and hence vanish for coefficients $s_{ab}$ with odd $a$. In the drizzled images the readout direction is up for the upper and down for the lower chip and the CTI trails occur in the opposite directions. This leads to a sign switch for coefficients with basis functions that are not symmetric about the $x$-axis (odd $b$), and we have taken this into account for $s_{03}$ in Figure~\ref{fig:cte}. We have detected (and modeled) a significant signature of CTI on the following stellar Shapelets coefficients: $s_{00}$, $s_{02}$, $s_{03}$, $s_{04}$, $s_{05}$, $s_{20}$, $s_{21}$, $s_{22}$, $s_{40}$, and $s_{60}$.

The correction of galaxy shapes for CTI again closely follows S10. Here we fit power-law corrections to the shear and (now in addition) flexion estimates as a function of galaxy flux, flux radius, sky background, time, and number of readout-transfers. Note that \citet{msl10} introduced a more advanced CTI correction scheme
operating directly on the pixel level. This is expected to yield higher precision, enabling for example the correction of the $s_{01}$ component, which cannot be estimated with our method due to its degeneracy with a simple shift in object position. However, we are confident that our correction scheme is sufficiently
accurate for the analysis presented here, in particular as potential residuals cancel to first order for the azimuthally averaged galaxy-galaxy lensing signal.

\subsection{Signal Computation}\label{app:cosmos_signal}
For the Navarro-Frenk-White (NFW) profile \citep*{nfw96}, the strength of the shear signal scales as
\begin{equation}
\gamma_{\mathrm{NFW}} \propto \frac{D_{l}D_{ls}}{D_{s}}
\end{equation}
where $D_{l}$, ${D_{s}}$ and $D_{ls}$ are the angular diameter distances to the lens, to the source, and between lens and source respectively \citep{wri00}. The flexion signals scale as
\begin{equation}
\mathcal{F}_{\mathrm{NFW}},\mathcal{G}_{\mathrm{NFW}} \propto  \frac{D^2_{l}D_{ls}}{D_{s}}
\end{equation}
\citepalias{bgr06}. We therefore weight the signals accordingly, scale them to a reference lens and source redshift and compute the weighted average in 25 logarithmic distance bins as follows:
\begin{equation}
\langle\gamma_t\rangle = \frac{\sum E_{\gamma_{t,i}}w_{\gamma_{t,i}}}{\sum w_{\gamma_{t,i}}}
\end{equation}\label{eq:weighted_average}
and similar for the flexions, with the shear estimator and weight
\begin{equation}
E_{\gamma_{t,i}} = \gamma_{t,i}\left(\frac{\eta_{i}}{\eta_{\mathrm{ref}}}\right)^{-1} \;\;\;\;\; w_{\gamma_{t,i}} = \frac{1}{\sigma_{\gamma,i}^2}\left(\frac{\eta_{i}}{\eta_{\mathrm{ref}}}\right)^2
\end{equation}\label{eq:shear_estimator}
where
\begin{equation}
\eta = \frac{D_{l}D_{ls}}{D_{s}}
\end{equation}\label{eq:lenseff}
is the geometric lensing efficiency and $\sigma^2_{\gamma,i} = \sigma^2_{\gamma,\mathrm{intr},i}+\sigma^2_{\gamma_{1},i}+\sigma^2_{\gamma_{2},i}$ is the error on the shape measurement with $\sigma_{\gamma,\mathrm{intr}}$ the intrinsic shear noise. By contrast we use the following F flexion estimator and weight:
\begin{equation}
E_{\mathcal{F}_{t,i}} = \mathcal{F}_{t,i}\left(\frac{D_{l,i}}{D_{l,\mathrm{ref}}}\frac{\eta_{i}}{\eta_{\mathrm{ref}}}\right)^{-1} \;\;\;\;\; w_{\mathcal{F}_{t,i}} = \frac{1}{\sigma^2_{\mathcal{F},i}}\left(\frac{D_{l,i}}{D_{l,\mathrm{ref}}}\frac{\eta_{i}}{\eta_{\mathrm{ref}}}\right)^2
\end{equation}\label{eq:fflex_estimator}
and similarly for the G flexion.

\section{High Redshift Results}\label{app:highz}
\begin{figure*}
\includegraphics[width=54mm,angle=270]{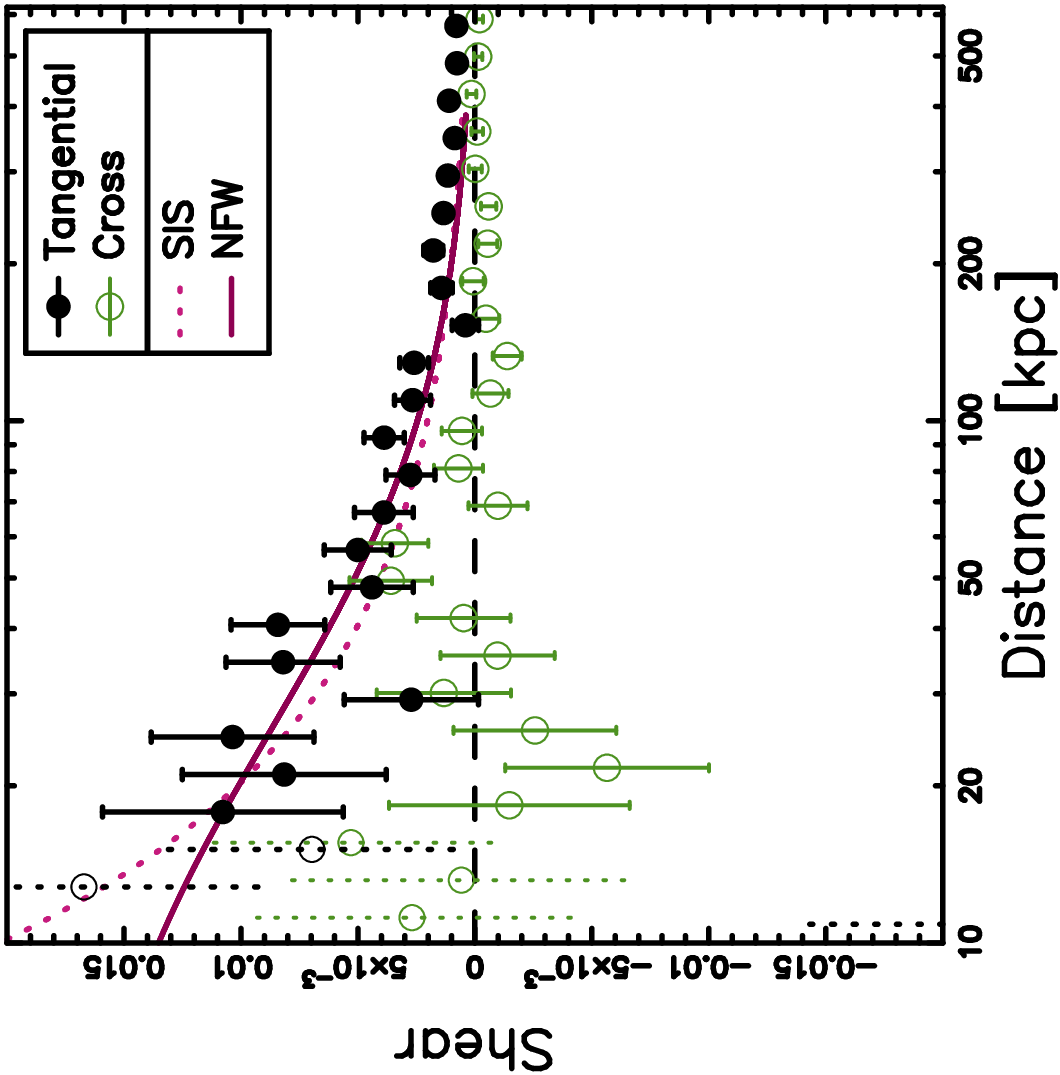}
\includegraphics[width=54mm,angle=270]{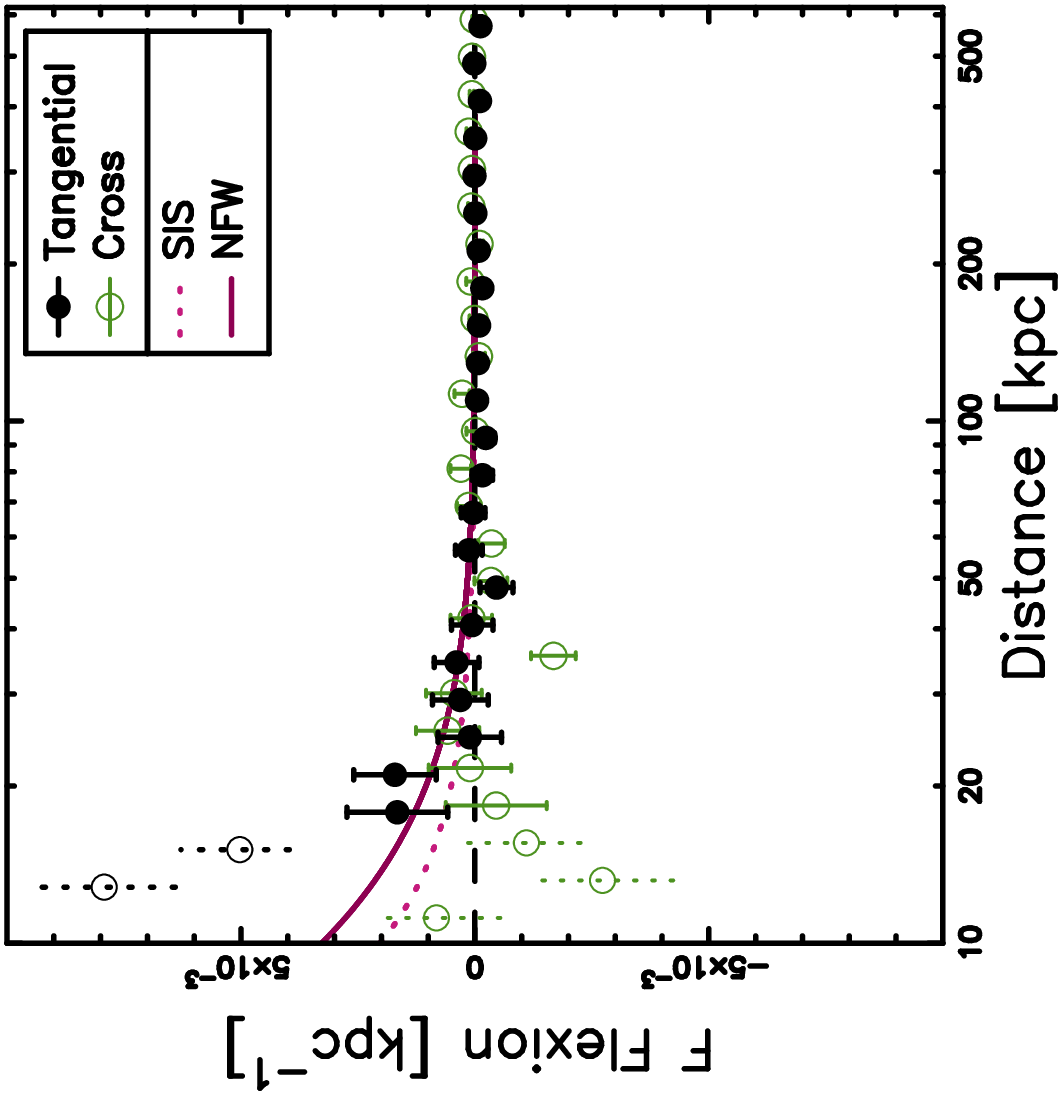}
\includegraphics[width=54mm,angle=270]{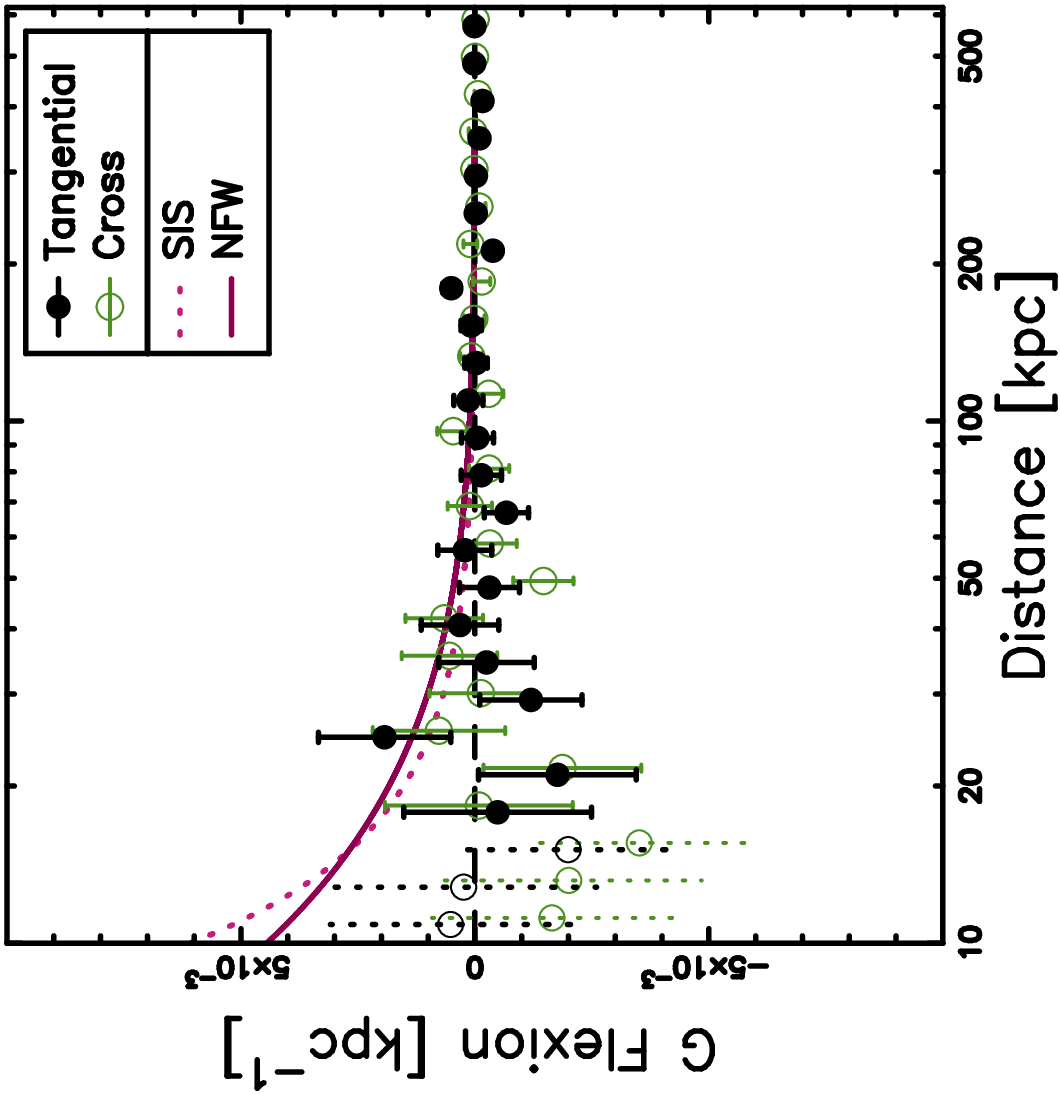}
\caption{ The galaxy-galaxy lensing results from running the MV pipeline on the COSMOS data, with \mbox{$n_\mathrm{max}=10$}. Black solid points represent the tangential signal and green circles represent the cross term. Empty circles with dotted error bars are bins that are too close to the lens on the sky. Please note that the SIS and NFW profiles have been fitted to the shear data and then {\em translated} into predictions for $\mathcal{F}$ and $\mathcal{G}$ curves.}
\label{app:fig:cosmos_highz}
\end{figure*}

As specified in the main paper, the lens catalogue we use has a redshift cut of $z<0.6$. This is to avoid having to go too close to the lens on the sky in order to see a flexion signal. Within an angular radius of $2\;\mathrm{arcsec}$ we have low confidence in the results; we are simply too close to the lensing galaxies and it becomes difficult to account for effects induced by the lens light. BOR corrects for light leakage at larger radii, but the correction is most likely incomplete very close to the lens due to deviations from a smooth S\'{e}rsic profile. For objects beyond our lens sample, the median redshift is close to 1.0. At this redshift the angular distance limit of $2\;\mathrm{arcsec}$ on the sky corresponds to a physical distance of about $17\;\mathrm{kpc}$. The F flexion falls off to low values already at about $20\;\mathrm{kpc}$ for a typical galaxy, so we are left with a very low signal within a narrow ring around the lens. Imposing the redshift cut of $z<0.6$ on lenses gives us a median lens redshift of $z=0.27$ at which the inner limit corresponds to $9\;\mathrm{kpc}$, leaving a wider distance interval in which we can investigate the F flexion signal.

In Figure~\ref{app:fig:cosmos_highz} we show the galaxy-galaxy signal for the high redshift sample, i.e.~for lenses with $z>0.6$. The bins that are within \mbox{2 arcsec} of the average lens in this sample, and which are most likely contaminated by lens light, are marked with dotted lines. The F flexion signal outside of this limit does agree well with the profile predicted by the shear, but falls off quickly.

\section{Comparison with KSB}\label{app:ksb}
\begin{figure}
\includegraphics[width=75mm,height=84mm,angle=270]{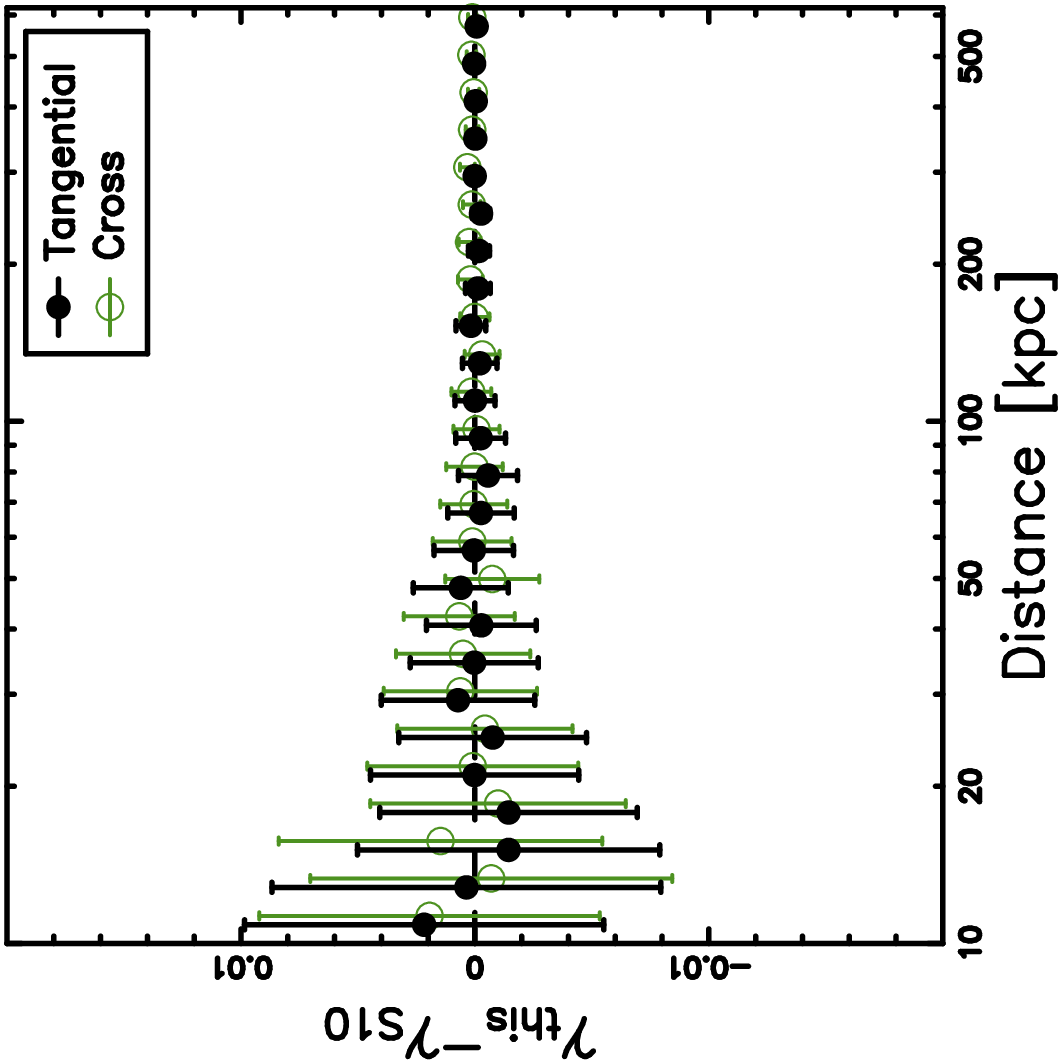}
\caption{ A comparison between the shears used in this paper and the ones used in \citetalias{shj10}. Black stars (green circles) show the difference between the tangential (cross) shear values in this paper and those produced by a KSB pipeline for \citetalias{shj10}.}
\label{app:fig:tim}
\end{figure}

We compare our galaxy-galaxy shear signal to the one we get using the shears from \citetalias{shj10}, using all the cuts normally applied in each analysis so that only common objects are used. The bias correction described in their paper is incorporated in their shears whilst our measurements have no correction applied. However, due to our S/N cut (see Appendix~\ref{app:cosmos_catalogues}) their correction is always less than 4.2\%.

As shown in Figure~\ref{app:fig:tim} the difference between the results from the two pipelines, KSB and Shapelets, is negligible. This provides an independent confirmation that the MV pipeline produces shears of as high a quality as the state-of-the-art weak lensing analysis presented in \citetalias{shj10}.

\label{lastpage}
\end{document}